# Adhesion of 2D MoS$_2$ to Graphite and Metal Substrates Measured by a Blister Test

Metehan Calis[1], David Lloyd[2], Narasimha Boddeti[3], and J. Scott Bunch[1,4*]

[1]Boston University, Department of Mechanical Engineering, Boston, MA 02215 USA

[2]Analog Garage, Analog Devices Inc, Boston, MA 02110 USA

[3]Washington State University, School of Mechanical and Materials Engineering, WA 99163 USA

[4]Boston University, Division of Materials Science and Engineering, Brookline, MA 02446 USA

*e-mail: bunch@bu.edu

## Abstract

Using a blister test, we measured the work of separation between MoS$_2$ membranes from metal, semiconductor, and graphite substrates. We found a work of separation ranging from 0.11 ± 0.05 J/m$^2$ for chromium to 0.39 ± 0.1 J/m$^2$ for graphite substrates. In addition, we measured the work of adhesion of MoS$_2$ membranes over these substrates and observed a dramatic difference between the work of separation and adhesion which we attribute to adhesion hysteresis. Due to the prominent role that adhesive forces play in the fabrication and functionality of devices made from 2D materials, an experimental determination of the work of separation and adhesion as provided here will help guide their development.

Two-dimensional(2D) materials[1] possess both remarkable mechanical properties[2–4] such as high tensile strength and impermeabilty[5,6] to gasses and extraordinary electrical and thermal properties such as high thermal conductivities,[7,8] large charge carrier mobilities,[9,10] and band gaps tunable by strain.[11] While initial research on 2D materials, focused on graphene, more recently this has been extended to 2D semiconductors such as transition metal dichalcogenides (TMDCs).[12] Combining different 2D materials to fabricate heterostructure devices[13,14] and using 2D single layers as building blocks,[15] more complex devices with advanced functionality are created.[16–18] Manufacturing of electrical and mechanical devices from 2D materials typically requires that the 2D layers are transferred to substrates and in the case of heterostructure devices the transfer involves stacking dissimilar 2D materials on top of each other.[19] The fabrication and performance, therefore, depend critically on the interfacial adhesion between the 2D materials and the surface it is in contact with. For this reason, understanding and controlling interfacial adhesion is of paramount importance.[20,21]

One of the most well-studied TMDCs is atomically thin MoS$_2$ which possesses exceptional mechanical strength,[22] flexibility,[23] high carrier mobility,[24] and a strain-tunable band gap.[25] While the adhesion of MoS$_2$ to various substrates such as other 2D materials and metals[21,26,27] has been previously studied, these studies have been limited to nanoscale areal regions, or centimeter size surfaces. Though both the nanoscale and macroscopic techniques offer unique insights into the delamination mechanics, these techniques may not be directly applicable to adhesion involved in delamination at the microscale where many devices from 2D materials operate.

Here, we use the constant-N pressurized blister test[28] to measure the work of separation and the work of adhesion of $MoS_2$ from metal, semiconductor, and graphite substrates. The constant-N pressurized blister test was previously used to determine Young's modulus,[29] adhesion energy,[30] and shear stress[31,32] of atomically thin 2D membranes. The ease of sample preparation and straightforward experimental configuration make this test ideally suited for such measurements at the microscale. Metals, semiconductor, and graphite are chosen as substrate materials because of their dissimilar electrical,[26-28,33,34] mechanical,[29,30] and optical[31,32] properties and their ubiquity as a substrate for 2D material devices.[35–38]

The blister test uses an applied pressure difference across a 2D membrane suspended over cylindrical microcavities.[29,30,39] For these experiments, we create etched cylindrical cavities on an oxidized silicon surface which is then coated with a metal or semiconductor such as gold (Au), chromium (Cr), titanium (Ti), germanium (Ge) or graphite. To fabricate the metal and Ge substrates, we pattern the silicon oxide ($SiO_x$)/silicon (Si) wafer with cylindrical microcavities using photolithography. Metal or Ge is evaporated over the substrate to form a thin film. Subsequently, chemical vapor deposition (CVD) grown MoS$_2$ flakes are transferred over the well to seal the microcavities (Figure 1a). To fabricate graphite wells, we start with the mechanical exfoliation of graphite over SiO$_x$/Si wafer. After this, microcavities are etched through the graphite, SiO$_x$, and Si (see the Supporting Information for further fabrication details). Using an optical microscope,[40] we select graphite layers whose thicknesses ranged from ~ 10 to 30 nm. A custom-made transfer station is then used to transfer MoS$_2$ flakes over the etched graphite wells thereby forming MoS$_2$ sealed microcavities (Figure 1b). Following the fabrication of both sets of devices, they are placed into a pressure chamber and charged to an input pressure ($p_0$) with argon gas. We wait ~ 48 h for gas to diffuse into the sealed microcavity such that the internal pressure ($p_{int}$) of the cavity and input pressure to reach the equilibrium ($p_0 = p_{int}$) before moving the devices to atmospheric pressure ($p_{ext} \equiv p_{atm} \approx$ 1 atm), $p_{int}$ is now larger than $p_{ext}$ ($\Delta p = p_{int} - p_{ext} > 0$) and results in the MoS$_2$ membrane bulging upward which we measure with an atomic force microscope (AFM) (Figure 1c). After each set of AFM measurements, the devices are placed back into the pressure chamber which is set to a higher $p_0$, and the test is repeated (Figure 2a). Initially, the membrane's radius stays constant ($a_0$), which is the radius of the well, until a critical pressure is reached. Once the internal pressure exceeds this critical pressure, the membrane delaminates from the surface, overcoming the adhesion forces creating a bulged blister. An AFM image of a bulged blister on a gold substrate showing the maximum deflection, $\delta$, and radius, $a$, is shown in Figure 1d. (see the Supporting Information for the other tested substrates).

Utilizing Hencky's solution[41,42] for the deflection of a fully clamped circular membranes subjected to a pressure load $\Delta p$, we obtain the relationship between $\Delta p$ and $\delta$ as

$$\Delta p = \frac{K(v) E_{2D} \delta^3}{a^4} \quad (1)$$

where $E_{2D}$ is the two-dimensional Young's modulus, and $K(v)$ is a constant which depends on the Poisson's ratio (see the Supporting Information). For MoS$_2$, we use $K(v = 0.29) = 3.54$.[30] Figure 2b shows $\Delta p$ vs $K(v) \delta^3/a^4$ for $a = a_0$. At each $\Delta p$, 3 AFM cross-sections separated by $\sim 10°$ degrees and passing through the center are used to determine and average $K(v) \delta^3/a^4$ along with a standard deviation. A linear fit to the data is used to determine $E_{2D}$ for each device (Figure 2c). The values of $E_{2D}$ are consistent with monolayer or bilayer MoS$_2$. The scattering of $E_{2D}$ values within the CVD-grown MoS$_2$ membranes may be caused by several factors such as the variations in the defects densities[43] and sulfur vacancies[44,45] from growth to growth, strain inhomogeneities which can be introduced during MoS$_2$ transfer over the wells,[46] and the possibility of bilayer flakes.

To determine the work of separation between the MoS$_2$ membrane and substrate, we use a previously developed free energy model[29,30,39] which assumes that the membrane, the membrane-substrate interface, the trapped gas, and the external atmosphere form an isothermal thermodynamic system whose free energy is given by:

$$F = \frac{(p_{int} - p_{ext}) V_b}{4} + \Gamma_{sep} \pi (a^2 - a_0^2) - p_o V_o \ln\left[\frac{V_o + V_b}{V_o}\right] + p_{ext} V_b \quad (2)$$

where $V_0$ is the initial volume of the microcavity and $V_b$ is bulge volume that is created after the membrane expansion (see the Supporting Information). The terms on the right-hand side are: (1) the strain energy of the membrane due to the pressure load, (2) the work of separation of the membrane from the substrate, where $\Gamma_{sep}$ is the separation energy per unit area, (3) isothermal work done by the trapped gas in the microcavity, and (4) work of the external pressure. When the critical pressure is exceeded, the membrane delaminates from the surface, and the blister expands until the free energy reaches its minimum value. Using eq 2, we find the local minima of the free energy of the delaminated configuration by taking the derivative with respect to $a$ and setting it to zero ($dF/da = 0$). Substituting eq 1 and the ideal gas law into the derivative yields

$$\Gamma_{sep} = \frac{5}{4} C K E_{2D} \left(\frac{\delta}{a}\right)^4 \qquad (3)$$

where $C(v = 0.29) = 0.522$.[30] Using eq 3, we can determine the work of separation for devices showing delamination. For each delaminated device using the measured $\delta$, $a$, and $E_{2D}$ of the corresponding devices (Figure 2c), we plot the mean values of $\Gamma_{sep}$ for the 7 devices that delaminated over the gold substrates, with error bars representing the standard deviation. The standard deviation was calculated by averaging the deflection and radius of the delaminated device along 3 cross sections separated by ~ 10 degrees. Some devices are measured at several increasing values of $p_0$ and undergo multiple delaminations to larger delaminated radii, $a$, and these are shown as identically colored data points (Figure 2d). We find an average $\Gamma_{sep} = 0.28 \pm 0.04$ J/m$^2$ (Figure 2d), which agrees closely with previous work (0.27 - 0.67 J/m$^2$).[47] The difference in the work of separation within each device may be caused by several factors, including local roughness differences[48] due to the etching process of the wells where the rim of the well may be rougher than other nearby surfaces, carbon based contamination on the surface,[12] or local differences in the chemical reactivity of the surface.[49] Measurements of strain-induced changes in photoluminescence (PL) of the MoS$_2$ membrane[25,50] confirm that the membrane remained well clamped to the substrate, thereby suggesting that Hencky's solution remains applicable throughout the delamination process (see the Supporting Information).  As shown in the previous studies, the strain gets extended outside the well area; however, incorporating the extended-strain into the regular Hencky solution has a minor effect (~ 1% change) on the work of separation.[30,31,51] We repeat the same experiment on the other metal and Ge substrates and find values lower than those of gold and SiO$_x$ with a noticeable difference in separation energies among the metals and Ge substrate tested (Figure 3). Possible mechanisms for the variations in adhesion can be electrostatic interactions[52,53] resulting in different binding affinities[54–56] or difference in surface roughness.  It is notable that gold has the highest work function (5.47 eV) within our tested substrates and does not form a native oxide such as Ge and the other metals tested.

We also measure the adhesion of MoS$_2$ on graphite, a semi-metal. The graphite surface is atomically smooth, assuming it is uncontaminated by surface residue and does not form a native oxide, though our graphite surfaces were slightly oxidized during processing (see the Supplementary Information). We perform the blister test on 44 graphite devices (see the Supporting Information for further details). Figure 4a shows the mean values of $\Gamma_{sep}$ for all the MoS$_2$/graphite devices. As can be seen, almost all of the devices undergo multiple delaminations. We find an average $\Gamma_{sep}$ = 0.39 ± 0.1 J/m$^2$ which is noticeably higher than the MoS$_2$ on gold devices ($\Gamma_{sep}$ = 0.28 ± 0.04 J/m$^2$) and higher than the MoS$_2$ on SiO$_x$ ($\Gamma_{sep}$ = 0.22 ± 0.01 J/m$^2$) (Figure 3). A clear illustration of the difference in $\Gamma_{sep}$ between MoS$_2$ on SiO$_x$ and MoS$_2$ over the graphite substrate is seen in Figure 4b-c where the MoS$_2$ is delaminated from the SiO$_x$ to a larger radius but remains pinned to its original radius over the graphite well. This is due to the larger Γ$_{sep}$ for MoS$_2$/graphite than MoS$_2$/SiO$_x$.

One possibility for the differences in Γ$_{sep}$ between MoS$_2$ and the metal or Ge substrates and graphite substrates is surface roughness. The surface roughness at the nanometer scale is critical to 2D material adhesion[48,57,58]. A freshly cleaved graphite surface should be atomically smooth and allow for better surface conformation[55,59,60] and perhaps better adhesion. Before the MoS$_2$ layers are transferred onto them, we find a graphite surface roughness of 0.21 ± 0.04 nm, which we attribute to photoresist contamination during processing. This roughness is similar to that of SiO$_x$ (0.18 ± 0.01 nm) and slightly less than the roughnesses of other metal or Ge surface (see in Table S2 in the Supporting Information for the details). The aluminum substrate had the largest surface roughness (3.8 ± 0.8 nm), and we were unable to transfer MoS$_2$ onto it, which we attribute to this large surface roughness, suggesting that surface roughness may play a role in our adhesion measurements. This suggests that the aluminum has a small work of adhesion. However, without higher resolution imaging of the surface such as that enabled by a scanning tunneling microscope, we are limited in the conclusion we can draw (see movies 1-4 in the Supporting Information for the transfer videos).

Another mechanism for adhesion differences is work function difference and chemical bonding between the 2D material and the substrate. Previously, it was shown that MoS$_2$ is more readily exfoliated onto freshly prepared gold substrates presumably due to a sulfur-gold bond that forms.[55] This could explain the larger adhesion we observe for gold over the other metals and Ge. Titanium, Cr, and Ge had similar surface roughness and correspondingly similar adhesion. They also all form a native oxide. Further studies are needed to confirm the exact role that both surface roughness, work function differences, and chemical bonding play in determining surface adhesion between 2D materials and their substrates.

In addition to measuring $\Gamma_{sep}$, we also measure the work of adhesion, $\Gamma_{adh}$. To do so, each device is monitored in the AFM under ambient conditions after delamination. Both the deflection and radius are continuously measured in the AFM over a period of 24 h to 10 days. As gas slowly leaks out of the well, the deflection decreases, but the radius stays pinned at the delaminated radius (Figure 5a). Once the deflection reaches a critical height, we see a gradual decrease in both the radius and the height of the blister.

In figure 5b, we plot $\delta$ vs $a$ for MoS$_2$ on gold and graphite samples undergoing both delamination and relamination. This difference in paths between delamination and relamination is attributed to adhesion hysteresis[61] similar to what we previously observed in MoS$_2$ on SiO$_x$[30,62,63] (see the Supporting Information for the other tested substrates).

Using a modified version of eq 3, we determine $\Gamma_{adh}$.[30] The onset of the relamination occurs at the critical deflection, $\delta_c$. The relamination process then takes place along a line where the slope is given by $\Gamma_{adh}$. The expression for the relamination energy can then be written as,

$$\Gamma_{adh} = \frac{5}{4} C K E_{2D} \left(\frac{\delta_c}{a_d}\right)^4 \tag{4}$$

where $\delta_c$ is the critical deflection and $a_d$ the radius at this deflection. Comparing Figure 3 and 5c, the energy that separates the membrane from the surface is considerably higher than the energy the membrane recovers upon relamination ($\Gamma_{sep} > \Gamma_{adh}$). We find $\Gamma_{adh}$ ranged from $\Gamma_{adh} = 0.057 \pm 0.008$ J/m² for graphite devices to $\Gamma_{adh} = 0.01 \pm 0.002$ J/m² for germanium devices (Figure 5c). (see the Supporting Information for complete data sets). This adhesion hysteresis was similar to what was previously observed in MoS$_2$ on SiO$_x$.[30]

The exact mechanism for the hysteresis isn't fully comprehended but one possible reason is that the membrane-substrate interface formed during the relamination is not equivalent that which is broken during delamination However, during pressurizing of the microcavities to $p_0$ the external pressure applied may alter the membrane-substrate interface by pushing the membrane onto the substrate and thereby increase the adhesive interactions prior to delamination measurements.[64–70]

In conclusion, we measured the work of separation of MoS$_2$ to graphite, Ge, Cr, Ti, and Au substrates using a blister test. We found $\Gamma_{sep}$ ranging from $0.08 \pm 0.03$ J/m² for Cr to 0.39 ± 0.1 J/m² for graphite substrates and a $\Gamma_{adh}$ value considerably lower than the $\Gamma_{sep}$ value. Our results suggest that both surface roughness and chemical interactions may play a role in surface adhesion of 2D materials but more research is needed to conclusively determine the exact contribution each makes. A measurement of both the work of separation and adhesion for a range of substrates as provided here is critical to guiding the future design of electrical and mechanical devices based on 2D materials due to the prominent role that these adhesive surface play in their fabrication and functionality. [18,71,72]

**Supporting information**

Supporting information covers the CVD growth and characterization, metal, Ge and graphite fabrication procedures, $MoS_2$ transfer method to substrates, graphite surface treatment, photoluminescence verification for clamping condition, data sets of graphite Young's modulus calculation, relamination data sets for metal, Ge and graphite substrates, surface roughness examination, and videos of $MoS_2$ transfer to $SiO_x$, Cr, Ti, and Al substrates.

**Acknowledgments:**

This work was funded by Ministry of National Education of Turkey under Graduate Education Scholarship (YLSY) program.

**References**

(1)  Novoselov, K. S.; Geim, A. K.; Morozov, S. V.; Jiang, D.; Zhang, Y.; Dubonos, S. V.; Grigorieva, I. V.; Firsov, A. A. Electric Field in Atomically Thin Carbon Films. *Science (80-. ).* **2004**, *306* (5696), 666–669. https://doi.org/10.1126/SCIENCE.1102896.

(2)  Bertolazzi, S.; Brivio, J.; Kis, A. Stretching and Breaking of Ultrathin MoS 2. *ACS Nano* **2011**, *5* (12), 9703–9709. https://doi.org/10.1021/nn203879f.

(3)  Lee, C.; Wei, X.; Kysar, J. W.; Hone, J. Measurement of the Elastic Properties and

Intrinsic Strength of Monolayer Graphene. *Science (80-. ).* **2008**, *321* (5887), 385–388. https://doi.org/10.1126/SCIENCE.1157996.

(4)  Geim, A. K. Graphene: Status and Prospects. *Science*. American Association for the Advancement of Science June 19, 2009, pp 1530–1534. https://doi.org/10.1126/science.1158877.

(5)  Sun, P. Z.; Yang, Q.; Kuang, W. J.; Stebunov, Y. V.; Xiong, W. Q.; Yu, J.; Nair, R. R.; Katsnelson, M. I.; Yuan, S. J.; Grigorieva, I. V.; Lozada-Hidalgo, M.; Wang, F. C.; Geim, A. K. Limits on Gas Impermeability of Graphene. *Nature* **2020**, *579* (7798), 229–232. https://doi.org/10.1038/s41586-020-2070-x.

(6)  Bunch, J. S.; Verbridge, S. S.; Alden, J. S.; Van Der Zande, A. M.; Parpia, J. M.; Craighead, H. G.; McEuen, P. L. Impermeable Atomic Membranes from Graphene Sheets. *Nano Lett.* **2008**, *8* (8), 2458–2462. https://doi.org/10.1021/NL801457B.

(7)  Lindsay, L.; Broido, D. A. Enhanced Thermal Conductivity and Isotope Effect in Single-Layer Hexagonal Boron Nitride. *Phys. Rev. B - Condens. Matter Mater. Phys.* **2011**, *84* (15), 155421. https://doi.org/10.1103/PHYSREVB.84.155421.

(8)  Zhang, G.; Zhang, Y. W. Thermoelectric Properties of Two-Dimensional Transition Metal Dichalcogenides. *J. Mater. Chem. C* **2017**, *5* (31), 7684–7698. https://doi.org/10.1039/C7TC01088E.

(9)  Bolotin, K. I.; Sikes, K. J.; Jiang, Z.; Klima, M.; Fudenberg, G.; Hone, J.; Kim, P.; Stormer, H. L. Ultrahigh Electron Mobility in Suspended Graphene. *Solid State Commun.* **2008**, *146* (9–10), 351–355. https://doi.org/10.1016/J.SSC.2008.02.024.

(10) Kim, S.; Konar, A.; Hwang, W. S.; Lee, J. H.; Lee, J.; Yang, J.; Jung, C.; Kim, H.; Yoo, J. B.; Choi, J. Y.; Jin, Y. W.; Lee, S. Y.; Jena, D.; Choi, W.; Kim, K. High-Mobility and Low-Power Thin-Film Transistors Based on Multilayer MoS2 Crystals. *Nat. Commun.* **2012**, *3* (1), 1–7. https://doi.org/10.1038/ncomms2018.

(11) Peng, Z.; Chen, X.; Fan, Y.; Srolovitz, D. J.; Lei, D. Strain Engineering of 2D Semiconductors and Graphene: From Strain Fields to Band-Structure Tuning and Photonic Applications. *Light Sci. Appl. 2020 91* **2020**, *9* (1), 1–25. https://doi.org/10.1038/s41377-020-00421-5.

(12) Choi, W.; Choudhary, N.; Han, G. H.; Park, J.; Akinwande, D.; Lee, Y. H. Recent Development of Two-Dimensional Transition Metal Dichalcogenides and Their Applications. *Mater. Today* **2017**, *20* (3), 116–130. https://doi.org/10.1016/J.MATTOD.2016.10.002.

(13) Liu, Y.; Huang, Y.; Duan, X. Van Der Waals Integration before and beyond Two-Dimensional Materials. *Nature* **2019**, *567* (7748), 323–333. https://doi.org/10.1038/s41586-019-1013-x.

(14) Buscema, M.; Island, J. O.; Groenendijk, D. J.; Blanter, S. I.; Steele, G. A.; Van Der Zant, H. S. J.; Castellanos-Gomez, A. Photocurrent Generation with Two-Dimensional van Der Waals Semiconductors. *Chem. Soc. Rev.* **2015**, *44* (11), 3691–3718. https://doi.org/10.1039/C5CS00106D.

(15) Geim, A. K.; Grigorieva, I. V. Van Der Waals Heterostructures. *Nature* **2013**, *499* (7459), 419–425. https://doi.org/10.1038/NATURE12385.

(16) Dean, C. R.; Young, A. F.; Meric, I.; Lee, C.; Wang, L.; Sorgenfrei, S.; Watanabe, K.; Taniguchi, T.; Kim, P.; Shepard, K. L.; Hone, J. Boron Nitride Substrates for High-Quality Graphene Electronics. *Nat. Nanotechnol.* **2010**, *5* (10), 722–726. https://doi.org/10.1038/nnano.2010.172.

(17) Ye, F.; Lee, J.; Feng, P. X. L. Atomic Layer $MoS_2$-Graphene van Der Waals Heterostructure Nanomechanical Resonators. *Nanoscale* **2017**, *9* (46), 18208–18215. https://doi.org/10.1039/C7NR04940D.

(18) Britnell, L.; Ribeiro, R. M.; Eckmann, A.; Jalil, R.; Belle, B. D.; Mishchenko, A.; Kim, Y. J.; Gorbachev, R. V.; Georgiou, T.; Morozov, S. V.; Grigorenko, A. N.; Geim, A. K.; Casiraghi, C.; Castro Neto, A. H.; Novoselov, K. S. Strong Light-Matter Interactions in Heterostructures of Atomically Thin Films. *Science (80-. ).* **2013**, *340* (6138), 1311–1314. https://doi.org/10.1126/SCIENCE.1235547.

(19) Ye, F.; Islam, A.; Zhang, T.; Feng, P. X. L. Ultrawide Frequency Tuning of Atomic Layer van Der Waals Heterostructure Electromechanical Resonators. *Nano Lett.* **2021**, *21* (13), 5508–5515. https://doi.org/10.1021/ACS.NANOLETT.1C00610.

(20) Liu, Y.; Xu, Z.; Zheng, Q. The Interlayer Shear Effect on Graphene Multilayer Resonators. *J. Mech. Phys. Solids* **2011**, *59* (8), 1613–1622. https://doi.org/10.1016/J.JMPS.2011.04.014.

(21) Li, B.; Yin, J.; Liu, X.; Wu, H.; Li, J.; Li, X.; Guo, W. Probing van Der Waals Interactions at Two-Dimensional Heterointerfaces. *Nat. Nanotechnol.* **2019**, *14* (6), 567–572. https://doi.org/10.1038/S41565-019-0405-2.

(22) Cooper, R. C.; Lee, C.; Marianetti, C. A.; Wei, X.; Hone, J.; Kysar, J. W. Nonlinear Elastic Behavior of Two-Dimensional Molybdenum Disulfide. *Phys. Rev. B - Condens. Matter Mater. Phys.* **2013**, *87* (3), 035423. https://doi.org/10.1103/PhysRevB.87.035423.

(23) Sharma, M.; Singh, A.; Singh, R. Monolayer $MoS_2$ Transferred on Arbitrary Substrates for Potential Use in Flexible Electronics. *ACS Appl. Nano Mater.* **2020**, *3* (5), 4445–4453. https://doi.org/10.1021/ACSANM.0C00551.

(24) Splendiani, A.; Sun, L.; Zhang, Y.; Li, T.; Kim, J.; Chim, C. Y.; Galli, G.; Wang, F. Emerging Photoluminescence in Monolayer $MoS_2$. *Nano Lett.* **2010**, *10* (4), 1271–1275. https://doi.org/10.1021/NL903868W.

(25) Lloyd, D.; Liu, X.; Christopher, J. W.; Cantley, L.; Wadehra, A.; Kim, B. L.; Goldberg, B. B.; Swan, A. K.; Bunch, J. S. Band Gap Engineering with Ultralarge

Biaxial Strains in Suspended Monolayer MoS2. *Nano Lett.* **2016**, *16* (9), 5836–5841. https://doi.org/10.1021/acs.nanolett.6b02615.

(26) Rokni, H.; Lu, W. Direct Measurements of Interfacial Adhesion in 2D Materials and van Der Waals Heterostructures in Ambient Air. *Nat. Commun.* **2020**, *11* (1), 1–14. https://doi.org/10.1038/s41467-020-19411-7.

(27) Fang, Z.; Li, X.; Shi, W.; Li, Z.; Guo, Y.; Chen, Q.; Peng, L.; Wei, X. Interlayer Binding Energy of Hexagonal MoS2as Determined by an in Situ Peeling-to-Fracture Method. *J. Phys. Chem. C* **2020**, *124* (42), 23419–23425. https://doi.org/10.1021/ACS.JPCC.0C06828.

(28) Wan, K. T.; Mai, Y. W. Fracture Mechanics of a New Blister Test with Stable Crack Growth. *Acta Metall. Mater.* **1995**, *43* (11), 4109–4115. https://doi.org/10.1016/0956-7151(95)00108-8.

(29) Boddeti, N. G.; Koenig, S. P.; Long, R.; Xiao, J.; Bunch, J. S.; Dunn, M. L. Mechanics of Adhered, Pressurized Graphene Blisters. *J. Appl. Mech. Trans. ASME* **2013**, *80* (4). https://doi.org/10.1115/1.4024255.

(30) Lloyd, D.; Liu, X.; Boddeti, N.; Cantley, L.; Long, R.; Dunn, M. L.; Bunch, J. S. Adhesion, Stiffness, and Instability in Atomically Thin MoS2 Bubbles. *Nano Lett.* **2017**, *17* (9), 5329–5334. https://doi.org/10.1021/acs.nanolett.7b01735.

(31) Wang, G.; Dai, Z.; Wang, Y.; Tan, P.; Liu, L.; Xu, Z.; Wei, Y.; Huang, R.; Zhang, Z. Measuring Interlayer Shear Stress in Bilayer Graphene. *Phys. Rev. Lett.* **2017**, *119* (3), 036101. https://doi.org/10.1103/PHYSREVLETT.119.036101.

(32) Kitt, A. L.; Qi, Z.; Rémi, S.; Park, H. S.; Swan, A. K.; Goldberg, B. B. How Graphene Slides: Measurement and Theory of Strain-Dependent Frictional Forces between Graphene and SiO2. *Nano Lett.* **2013**, *13* (6), 2605–2610. https://doi.org/10.1021/nl4007112.

(33) Santiago, Y.; Cabrera -, C. R.; Garcia, B.; Roy, F.; Bélanger, D.; Zhang, Z.; Chen, K.; Zhao, Q.; Huang, C.; Jin, Y.; Wang, W.; Tang, L.; Song, C.; Xiu, F. Manganese and Chromium Doping in Atomically Thin MoS2*. *J. Semicond.* **2017**, *38* (3), 033004. https://doi.org/10.1088/1674-4926/38/3/033004.

(34) Andzane, J.; Petkov, N.; Livshits, A. I.; Boland, J. J.; Holmes, J. D.; Erts, D. Two-Terminal Nanoelectromechanical Devices Based on Germanium Nanowires. *Nano Lett.* **2009**, *9* (5), 1824–1829. https://doi.org/10.1021/NL8037807.

(35) Zhao, X.; Yang, S.; Sun, Z.; Cui, N.; Zhao, P.; Tang, Q.; Tong, Y.; Liu, Y. Enhancing the Intrinsic Stretchability of Micropatterned Gold Film by Covalent Linkage of Carbon Nanotubes for Wearable Electronics. *ACS Appl. Electron. Mater.* **2019**, *1* (7), 1295–1303. https://doi.org/10.1021/ACSAELM.9B00243.

(36) Abbasi, N. M.; Xiao, Y.; Zhang, L.; Peng, L.; Duo, Y.; Wang, L.; Yin, P.; Ge, Y.; Zhu, H.; Zhang, B.; Xie, N.; Duan, Y.; Wang, B.; Zhang, H. Heterostructures of Titanium-Based MXenes in Energy Conversion and Storage Devices. *J. Mater. Chem. C* **2021**, *9* (27), 8395–8465. https://doi.org/10.1039/D1TC00327E.

(37) Lemme, M. C.; Wagner, S.; Lee, K.; Fan, X.; Verbiest, G. J.; Wittmann, S.; Lukas, S.; Dolleman, R. J.; Niklaus, F.; van der Zant, H. S. J.; Duesberg, G. S.; Steeneken, P. G. Nanoelectromechanical Sensors Based on Suspended 2D Materials. *Research* **2020**, *2020*, 1–25. https://doi.org/10.34133/2020/8748602.

(38) Frisenda, R.; Niu, Y.; Gant, P.; Muñoz, M.; Castellanos-Gomez, A. Naturally Occurring van Der Waals Materials. *npj 2D Mater. Appl. 2020 41* **2020**, *4* (1), 1–13. https://doi.org/10.1038/s41699-020-00172-2.

(39) Koenig, S. P.; Boddeti, N. G.; Dunn, M. L.; Bunch, J. S. Ultrastrong Adhesion of Graphene Membranes. *Nat. Nanotechnol.* **2011**, *6* (9), 543–546. https://doi.org/10.1038/nnano.2011.123.

(40) Li, H.; Wu, J.; Huang, X.; Lu, G.; Yang, J.; Lu, X.; Xiong, Q.; Zhang, H. Rapid and Reliable Thickness Identification of Two-Dimensional Nanosheets Using Optical Microscopy. *ACS Nano* **2013**, *7* (11), 10344–10353. https://doi.org/10.1021/nn4047474.

(41) Fichter, W. B. Some Solutions for the Large Deflections of Uniformly Loaded Circular Membranes. *NASA Tech. Pap.* **1997**, *3658*, 1–24.

(42) Hencky, H. Ueber Den Spannungszustand in Kreisrunden Platten Mit Verschwindender Biegungssteifigkeit. *Zeitschrift für Math. und Phys.* **1915**, *63*, 311–317.

(43) Zandiatashbar, A.; Lee, G. H.; An, S. J.; Lee, S.; Mathew, N.; Terrones, M.; Hayashi, T.; Picu, C. R.; Hone, J.; Koratkar, N. Effect of Defects on the Intrinsic Strength and Stiffness of Graphene. *Nat. Commun. 2014 51* **2014**, *5* (1), 1–9. https://doi.org/10.1038/ncomms4186.

(44) Hong, J.; Hu, Z.; Probert, M.; Li, K.; Lv, D.; Yang, X.; Gu, L.; Mao, N.; Feng, Q.; Xie, L.; Zhang, J.; Wu, D.; Zhang, Z.; Jin, C.; Ji, W.; Zhang, X.; Yuan, J.; Zhang, Z. Exploring Atomic Defects in Molybdenum Disulphide Monolayers. *Nat. Commun. 2015 61* **2015**, *6* (1), 1–8. https://doi.org/10.1038/ncomms7293.

(45) Gan, Y.; Zhao, H. Chirality Effect of Mechanical and Electronic Properties of Monolayer MoS2 with Vacancies. *Phys. Lett. A* **2014**, *378* (38–39), 2910–2914. https://doi.org/10.1016/J.PHYSLETA.2014.08.008.

(46) Arjmandi-Tash, H.; Allain, A.; Han, Z.; -, al; Chen, C.; Avila, J.; Liu, B.; Pavlou, C.; Wang, Z.; Cang, Y.; Galiotis, C.; Fytas, G. Determination of the Elastic Moduli of CVD Graphene by Probing Graphene/Polymer Bragg Stacks. *2D Mater.* **2021**, *8* (3), 035040. https://doi.org/10.1088/2053-1583/ABFEDB.

(47) Fang, Z.; Dai, Z.; Wang, B.; Tian, Z.; Yu, C.; Chen, Q.; Wei, X. Pull-to-Peel of Two-Dimensional Materials for the Simultaneous Determination of Elasticity and Adhesion. *Nano Lett.* **2022**, *03*, 23. https://doi.org/https://pubs.acs.org/doi/full/10.1021/acs.nanolett.2c03145.

(48) Boddeti, N. G.; Long, R.; Dunn, M. L. Adhesion Mechanics of Graphene on Textured Substrates. *Int. J. Solids Struct.* **2016**, *97–98*, 56–74. https://doi.org/10.1016/J.IJSOLSTR.2016.07.043.

(49) Liu, X.; Boddeti, N. G.; Szpunar, M. R.; Wang, L.; Rodriguez, M. A.; Long, R.; Xiao, J.; Dunn, M. L.; Bunch, J. S. Observation of Pull-in Instability in Graphene Membranes under Interfacial Forces. *Nano Lett.* **2013**, *13* (5), 2309–2313. https://doi.org/10.1021/nl401180t.

(50) Yang, R.; Lee, J.; Ghosh, S.; Tang, H.; Sankaran, R. M.; Zorman, C. A.; Feng, P. X. L. Tuning Optical Signatures of Single- and Few-Layer MoS2 by Blown-Bubble Bulge Straining up to Fracture. *Nano Lett.* **2017**, *17* (8), 4568–4575. https://doi.org/10.1021/ACS.NANOLETT.7B00730.

(51) Ma, Y.; Wang, G.; Chen, Y.; Long, D.; Guan, Y.; Liu, L.; Zhang, Z. Extended Hencky Solution for the Blister Test of Nanomembrane. *Extrem. Mech. Lett.* **2018**, *22*, 69–78. https://doi.org/10.1016/J.EML.2018.05.006.

(52) Vernov, A.; Steele, W. A. The Electrostatic Field at a Graphite Surface and Its Effect on Molecule-Solid Interactions. *Langmuir* **1992**, *8* (1), 155–159. https://doi.org/10.1021/la00037a029.

(53) Speake, C. C.; Trenkel, C. Forces between Conducting Surfaces Due to Spatial Variations of Surface Potential. *Phys. Rev. Lett.* **2003**, *90* (16), 4. https://doi.org/10.1103/PhysRevLett.90.160403.

(54) Velický, M.; Rodriguez, A.; Bouša, M.; Krayev, A. V.; Vondráček, M.; Honolka, J.; Ahmadi, M.; Donnelly, G. E.; Huang, F.; Abruña, H. D.; Novoselov, K. S.; Frank, O. Strain and Charge Doping Fingerprints of the Strong Interaction between Monolayer MoS2 and Gold. *J. Phys. Chem. Lett.* **2020**, *11* (15), 6112–6118. https://doi.org/10.1021/ACS.JPCLETT.0C01287.

(55) Velický, M.; Donnelly, G. E.; Hendren, W. R.; McFarland, S.; Scullion, D.; DeBenedetti, W. J. I.; Correa, G. C.; Han, Y.; Wain, A. J.; Hines, M. A.; Muller, D. A.; Novoselov, K. S.; Abruńa, H. D.; Bowman, R. M.; Santos, E. J. G.; Huang, F. Mechanism of Gold-Assisted Exfoliation of Centimeter-Sized Transition-Metal Dichalcogenide Monolayers. *ACS Nano* **2018**, *12* (10), 10463–10472. https://doi.org/10.1021/ACSNANO.8B06101.

(56) Chen, X.; Tian, F.; Persson, C.; Duan, W.; Chen, N. X. Interlayer Interactions in Graphites. *Sci. Rep.* **2013**, *3* (1), 1–5. https://doi.org/10.1038/srep03046.

(57) Aitken, Z. H.; Huang, R. Effects of Mismatch Strain and Substrate Surface Corrugation on Morphology of Supported Monolayer Graphene. *J. Appl. Phys.* **2010**, *107* (12), 123531. https://doi.org/10.1063/1.3437642.

(58) Gao, W.; Huang, R. Effect of Surface Roughness on Adhesion of Graphene Membranes. *J. Phys. D. Appl. Phys.* **2011**, *44* (45), 452001. https://doi.org/10.1088/0022-3727/44/45/452001.

(59) Xin, H.; Borduin, R.; Jiang, W.; Liechti, K. M.; Li, W. Adhesion Energy of As-Grown Graphene on Copper Foil with a Blister Test. *Carbon N. Y.* **2017**, *123*, 243–249. https://doi.org/10.1016/J.CARBON.2017.07.053.

(60) Gao, W.; Huang, R. Effect of Surface Roughness on Adhesion of Graphene Membranes. *J. Phys. D. Appl. Phys.* **2011**, *44* (45), 452001. https://doi.org/10.1088/0022-3727/44/45/452001.

(61) Kesari, H.; Doll, J. C.; Pruitt, B. L.; Cai, W.; Lew, A. J. Role of Surface Roughness in Hysteresis during Adhesive Elastic Contact. *Philos. Mag. Lett.* **2010**, *90* (12), 891–902. https://doi.org/10.1080/09500839.2010.521204.

(62) Miskin, M. Z.; Sun, C.; Cohen, I.; Dichtel, W. R.; McEuen, P. L. Measuring and Manipulating the Adhesion of Graphene. *Nano Lett.* **2018**, *18* (1), 449–454. https://doi.org/10.1021/ACS.NANOLETT.7B04370.

(63) Chen, Y. L.; Helm, C. A.; Israelachvili, J. N. Molecular Mechanisms Associated with Adhesion and Contact Angle Hysteresis of Monolayer Surfaces. *J. Phys. Chem.* **1991**, *95* (26), 10736–10747. https://doi.org/10.1021/J100179A041.

(64) Girard-Reydet, E.; Oslanec, R.; Whitten, P.; Brown, H. R. Effects of Contact Time and Polarity Level on Adhesion between Interacting Surfaces. *Langmuir* **2004**, *20* (3), 708–713. https://doi.org/10.1021/LA034976B.

(65) Maugis, D.; Barquins, M. Fracture Mechanics and the Adherence of Viscoelastic Bodies. *J. Phys. D. Appl. Phys.* **1978**, *11* (14), 1989. https://doi.org/10.1088/0022-3727/11/14/011.

(66) Suk, J. W.; Na, S. R.; Stromberg, R. J.; Stauffer, D.; Lee, J.; Ruoff, R. S.; Liechti, K. M. Probing the Adhesion Interactions of Graphene on Silicon Oxide by Nanoindentation. *Carbon N. Y.* **2016**, *103*, 63–72. https://doi.org/10.1016/J.CARBON.2016.02.079.

(67) Kim, S.; Choi, G. Y.; Ulman, A.; Fleischer, C. Effect of Chemical Functionality on Adhesion Hysteresis. *Langmuir* **1997**, *13* (25), 6850–6856. https://doi.org/10.1021/LA970649Q.

(68) Makkonen, L. A Thermodynamic Model of Contact Angle Hysteresis. *J. Chem. Phys.* **2017**, *147* (6), 064703. https://doi.org/10.1063/1.4996912.

(69) Kang, J.; Sahin, H.; Peeters, F. M. Mechanical Properties of Monolayer Sulphides: A Comparative Study between MoS2, HfS2 and TiS3. *Phys. Chem. Chem. Phys.* **2015**, *17* (41), 27742–27749. https://doi.org/10.1039/C5CP04576B.

(70) Qiao, S.; Gratadour, J. B.; Wang, L.; Lu, N. Conformability of a Thin Elastic Membrane Laminated on a Rigid Substrate With Corrugated Surface. *IEEE Trans.*

*Components, Packag. Manuf. Technol.* **2015**, *5* (9), 1237–1243. https://doi.org/10.1109/TCPMT.2015.2453319.

(71) Furchi, M. M.; Höller, F.; Dobusch, L.; Polyushkin, D. K.; Schuler, S.; Mueller, T. Device Physics of van Der Waals Heterojunction Solar Cells. *npj 2D Mater. Appl. 2018 21* **2018**, *2* (1), 1–7. https://doi.org/10.1038/s41699-018-0049-3.

(72) Bediako, D. K.; Rezaee, M.; Yoo, H.; Larson, D. T.; Zhao, S. Y. F.; Taniguchi, T.; Watanabe, K.; Brower-Thomas, T. L.; Kaxiras, E.; Kim, P. Heterointerface Effects in the Electrointercalation of van Der Waals Heterostructures. *Nature* **2018**, *558* (7710), 425–429. https://doi.org/10.1038/s41586-018-0205-0.

(73) Liu, K.; Yan, Q.; Chen, M.; Fan, W.; Sun, Y.; Suh, J.; Fu, D.; Lee, S.; Zhou, J.; Tongay, S.; Ji, J.; Neaton, J. B.; Wu, J. Elastic Properties of Chemical-Vapor-Deposited Monolayer MoS2, WS2, and Their Bilayer Heterostructures. *Nano Lett.* **2014**, *14* (9), 5097–5103. https://doi.org/10.1021/nl501793a.

(74) Castellanos-Gomez, A.; Poot, M.; Steele, G. A.; Van Der Zant, H. S. J.; Agraït, N.; Rubio-Bollinger, G. Elastic Properties of Freely Suspended MoS2 Nanosheets. *Adv. Mater.* **2012**, *24* (6), 772–775. https://doi.org/10.1002/adma.201103965.

**Figure 1**

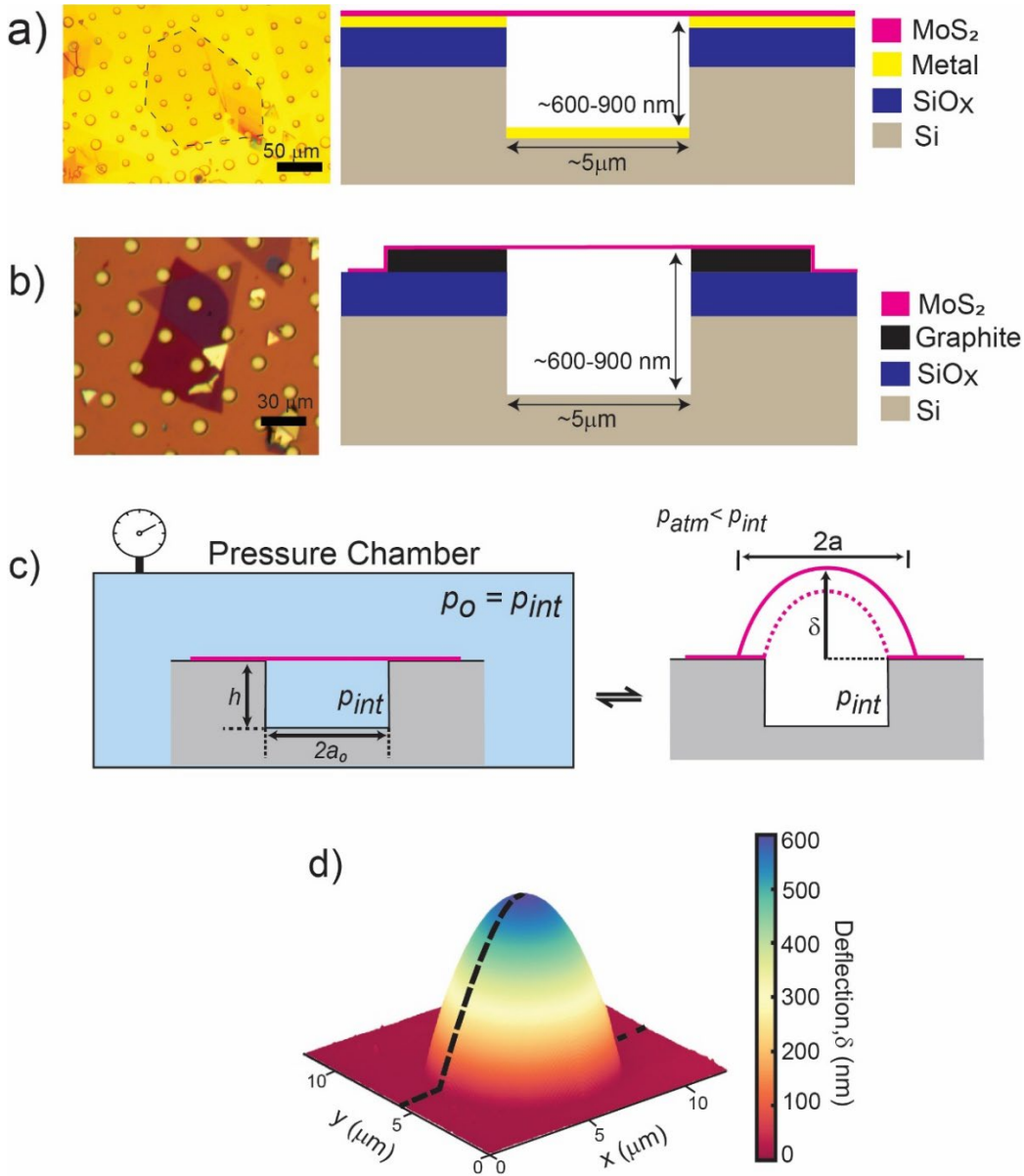

**Figure 1. a)** (left) Optical image of the MoS$_2$ over the gold substrate. The dashed lines show the boundary of the MoS$_2$ flakes. (right) Schematic image of the device with the gold substrate. **b)** (left) Optical image of the MoS$_2$ over the graphite substrate and (right) the schematic of the device. **c)** Schematic illustration of the experimental procedure. Devices are placed into a pressure chamber and kept until the $p_o = p_{int}$. When the devices are taken out, the membrane bulges up due to $p_{int} > p_{ext}$ ($\approx p_{atm}$). This process is then repeated until the delamination. **d)** AFM image of the blister. The dashed line represents a line cut passing through the center of the blister from which the maximum deflection and radius are obtained.

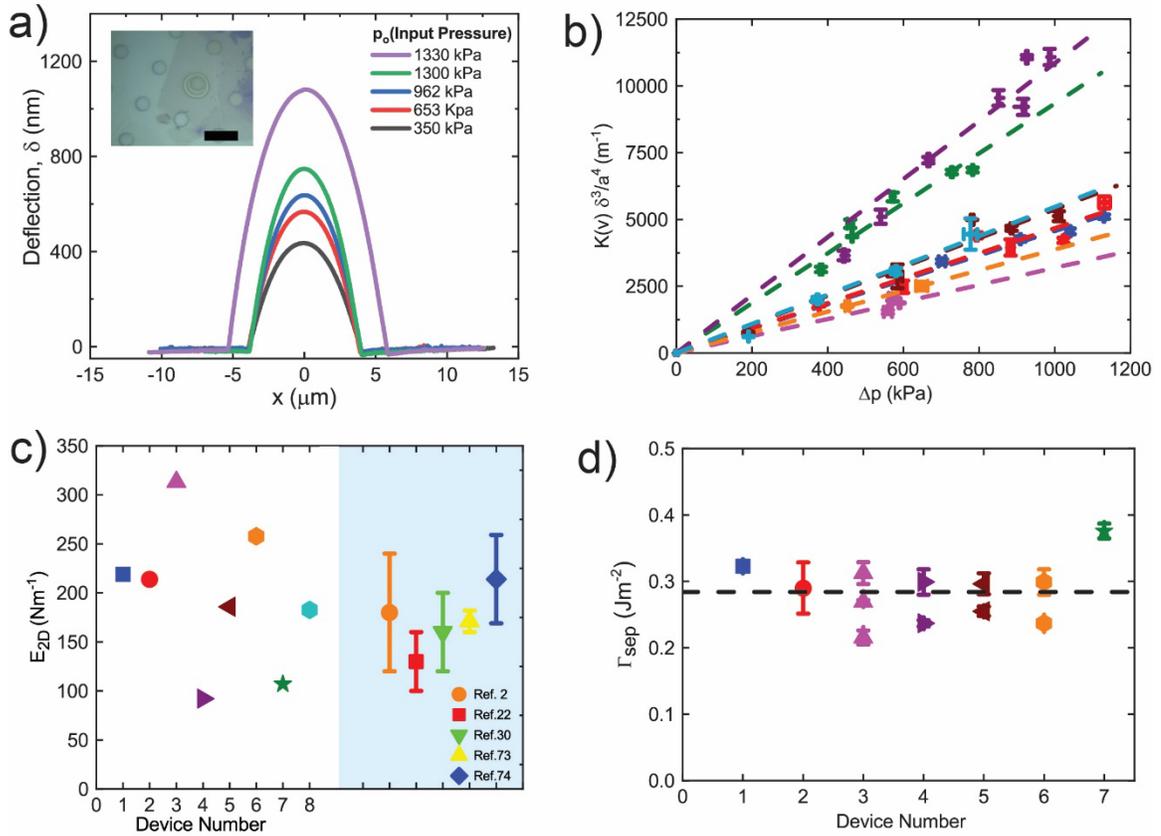

**Figure 2. MoS$_2$ on Gold substrate: a)** AFM cross sections of the delaminated devices at various input pressure. Inset: optical image of the delaminated device over the gold substrate (Scale bar is 15μm). **b)** $K(v)\delta^3/a^4$ vs $\Delta p$ for CVD-grown MoS$_2$ membranes. Dashed lines are linear fits used to calculate E$_{2D}$ for each device. Different colors/symbols represent the different devices. **c)** E$_{2D}$ for each device in (b). Reference values are plotted for comparison. [2,22,30,73,74] **d)** The corresponding E$_{2D}$ for each device are used in the calculation of its work of separation. Several devices were subjected to multiple delaminations from the surface (one device got broken before delamination). Data points are the mean values and the error bars represent the standard deviations. The dashed line is the average of all devices.

**Figure 3**

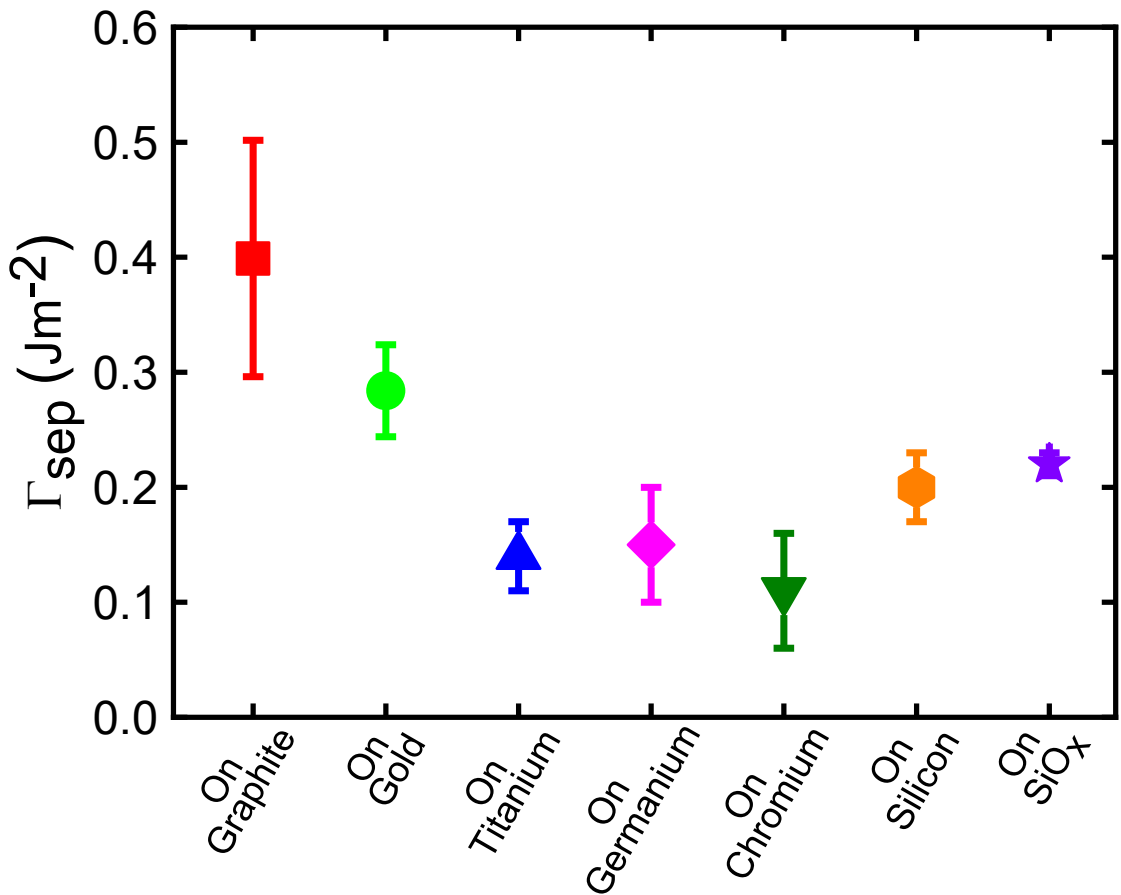

**Figure 3.** Comparison of work of separation between the graphite, Cr, Ti, Au, Ge, $SiO_x$, and Si substrates. Data points are mean values, and error bars represent the standard deviations.

**Figure 4**

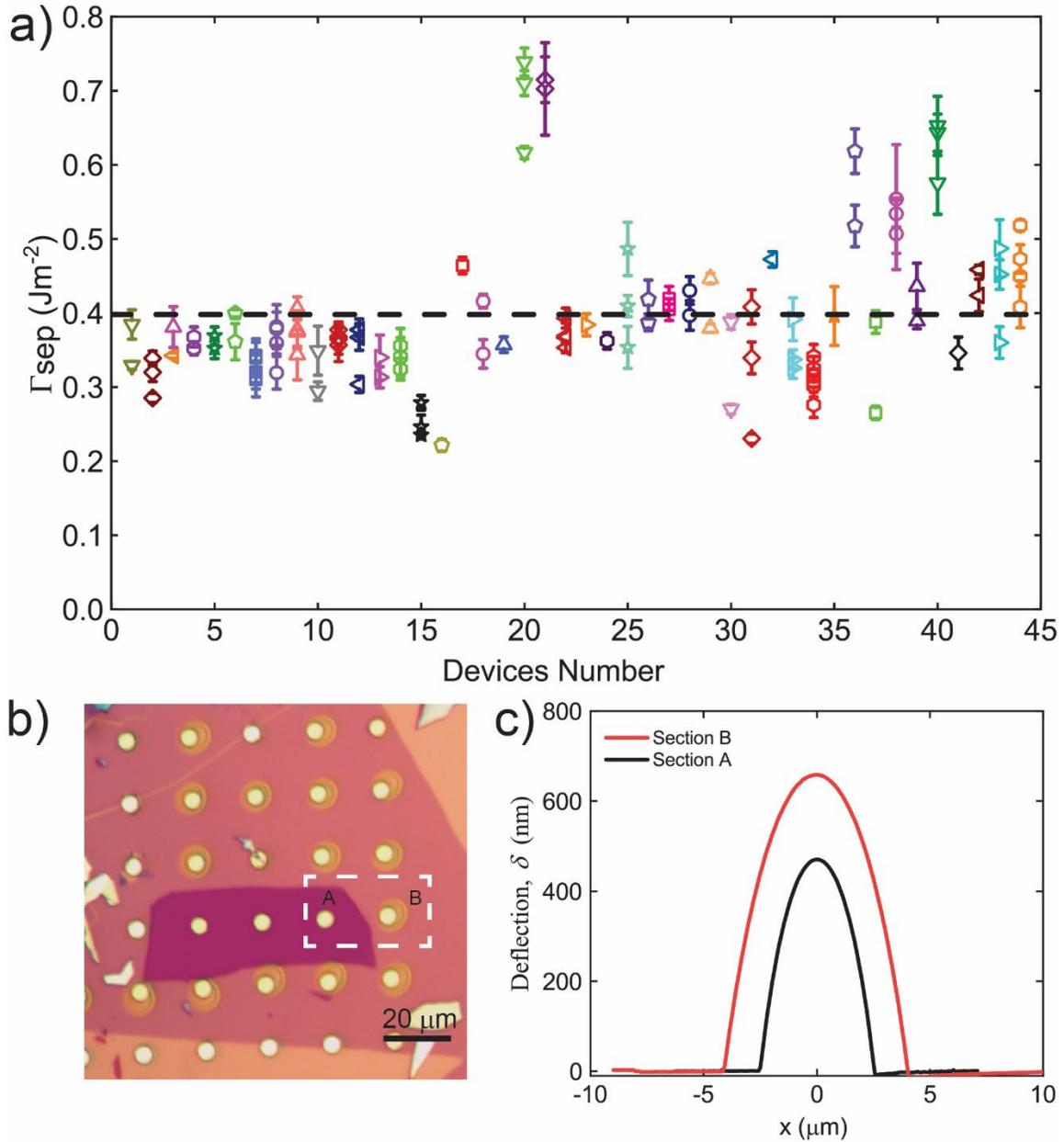

**Figure 4. a)** Work of separation for the graphite devices. Most of the devices were subjected to multiple delaminations from the surface. Data points are mean values and the error bars represent the standard deviations. The dashed line is the average of all devices. **b)** Optical image of the $MoS_2$ over the graphite. **c)** Cross sections of the devices labeled with the dashed line in (b) at the same input pressure. The membrane over the $SiO_x$ shows earlier delamination due to a lower work of separation.

**Figure 5**

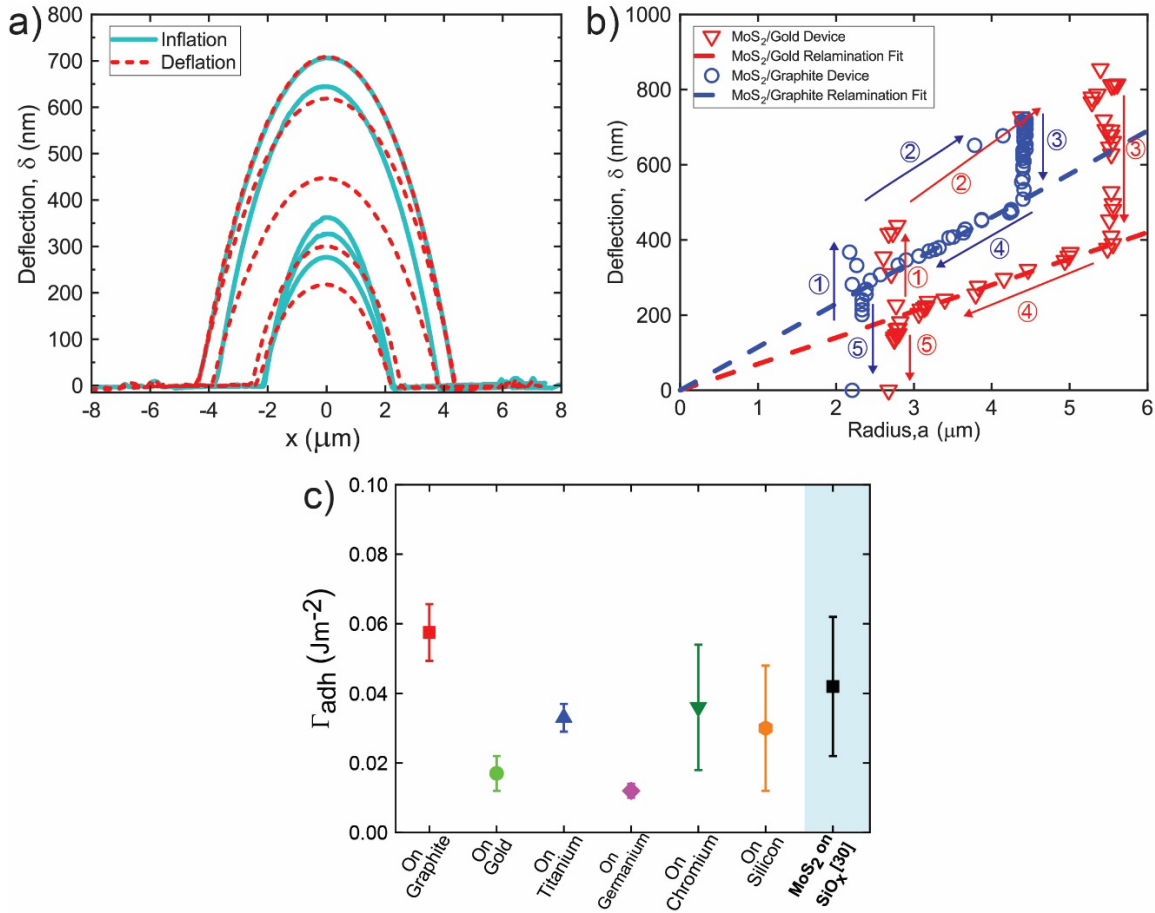

**Figure 5. a)** AFM cross section of the device during inflation and deflation **b)** $\delta$ and $a$ during the inflation and deflation. Gold and a graphite substrates are shown (see the Supporting Information for more data). Arrows show various regimes: (1) inflation, (2) separation, (3) deflation at a pinned radius (4) relamination, and (5) deflation at the well's radius. Dashed lines are the linear fits for the calculation of the work of adhesion. **c)** Comparison of work of adhesion between graphite, Au, Ti, Ge, Cr, and Si substrates. Data points are mean values, and error bars represent the standard deviation.

Supporting Information

# Adhesion of 2D MoS$_2$ to Graphite and Metal Substrates Measured by a Blister Test

Metehan Calis[1], David Lloyd[2], Narasimha Boddeti[3], and J. Scott Bunch[1,4*]

[1]Boston University, Department of Mechanical Engineering, Boston, MA 02215 USA

[2]Analog Garage, Analog Devices Inc, Boston, MA 02110 USA

[3]Washington State University, School of Mechanical and Materials Engineering, WA 99163 USA

[4]Boston University, Division of Materials Science and Engineering, Brookline, MA 02446 USA

*e-mail: bunch@bu.edu

# 1. Growth and Characterization

We used chemical vapor deposition (CVD) to grow monolayer $MoS_2$ flakes[1,2] (Fig. S1a). Initially, $MoS_2$ powder (Thermo Fisher Scientific, Molybdenum (IV) sulfide, 98%) is placed into the middle of the furnace inside an aluminum oxide crucible. A $SiO_x$ wafer, which is cleaned with acetone, isopropyl alcohol (IPA), and deionized (DI) water and then exposed to ultraviolet (UV) for 5 minutes, is placed into a cooler region downstream of the $MoS_2$ powder. Before starting the growth, the furnace is put under vacuum and purged with Argon (200 sccm) to remove air (specifically $O_2$). We then introduce 60 sccm Ar, 0.06 sccm $O_2$, and 1.8 sccm $H_2$ inside the tube. The growth process consists of three steps: (i) heating up to 900 °C for 15 minutes, (ii) holding at 900 °C for 15 minutes, and (iii) cooling the furnace to room temperature. During the growth, $MoS_2$ powder sublimates and is carried downstream by Ar gas where it condenses onto the cooler $SiO_x$ substrate.

After the growth, we observe numerous monolayer flakes in the optical microscope with ~ 100 μm side length. We deduce the approximate number of layers by using optical contrast[3] and confirm this with both Raman spectroscopy (Fig. S1b) and photoluminescence (PL) spectroscopy (Fig. S1c). Both Raman and PL Spectroscopy were conducted in a Renishaw Raman InVia microscope using a 532 nm laser beam with 1200 l/mm gratings. In Raman spectroscopy, we obtained two prominent resonances[4], the in-plane ($E^1_{2g}$) and out-of-plane ($A_{1g}$) vibrations located at 385 cm$^{-1}$ and 405 cm$^{-1}$ respectively, consistent with monolayer flakes. PL measurements[5] show a peak at 1.9 eV (A exciton) and the absence of an indirect peak, both indicative of monolayer $MoS_2$.

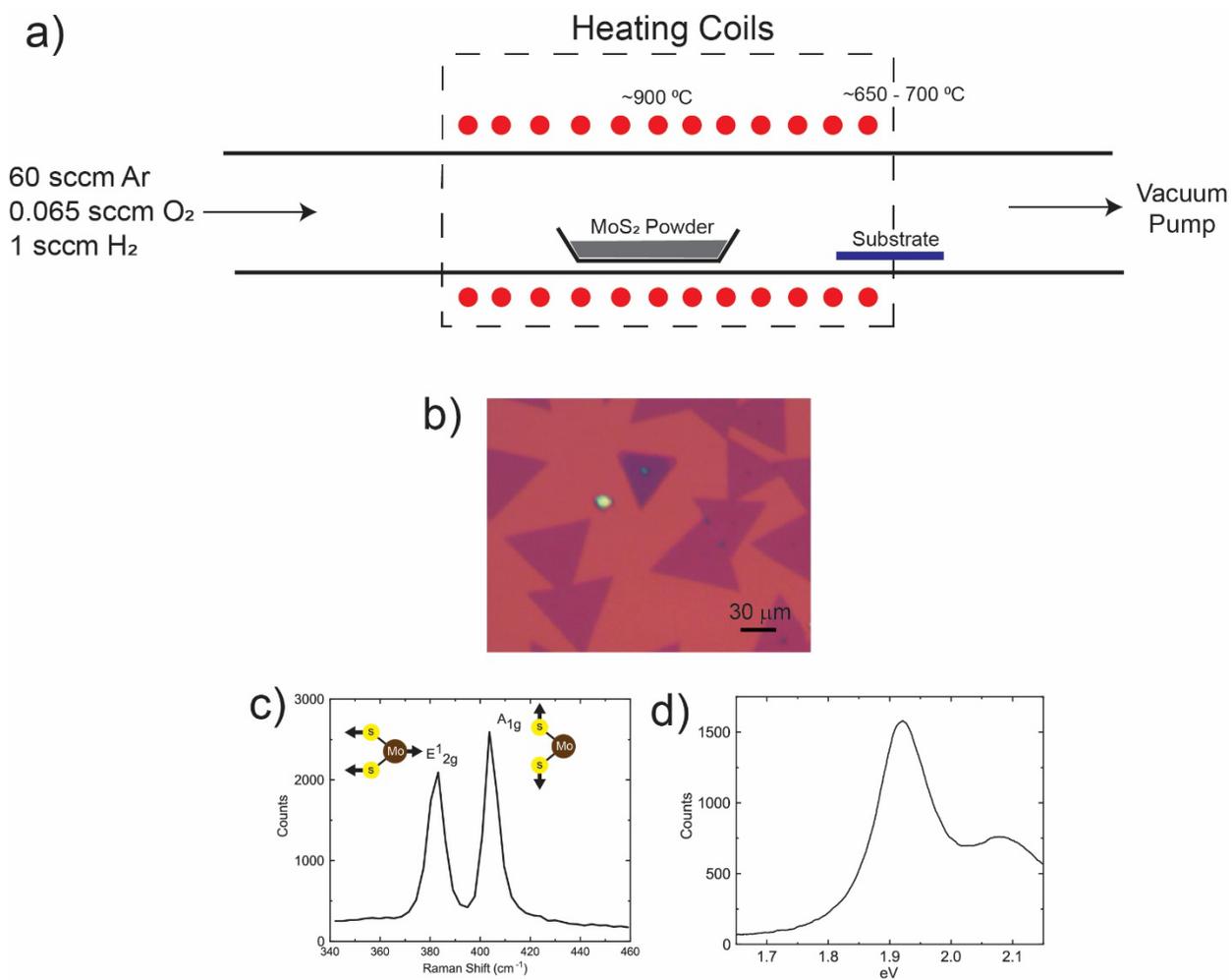

**Figure S1. a)** Schematics of the positions of the crucible and substrate in the furnace. **b)** Monolayer MoS$_2$ flakes on the SiO$_x$ substrate after the growth. **c)** Raman spectrum (Inset: Schematic of vibrational modes, E$^1_{2g}$ and A$_{1g}$). **d)** PL spectrum.

## 2. Metal Substrate Fabrication

Here we describe briefly the steps of the fabrication of the metal substrates. First, we clean the $SiO_x$ chips with acetone, IPA, and DI water. The surface of the $SiO_x$ is spin-coated with a photoresist solution of S1818 at 2500 rpm, and the substrate is put on a hot plate at 115 °C for 1 minute. We use a contact aligner with a mask with 5 μm diameter circles on it to create the photoresist pattern. The spin-coated chips are exposed to UV for 20 seconds with 8 mW power. The parts which are exposed to UV are removed with MF-319 Developer and then $SiO_x$ and Si are etched by Reactive-Ion-Etching (RIE). We use 3.1 sccm $O_2$ and 25 sccm $CF_4$ at 100 mTorr pressure and 150 W power for 13 minutes of etching. This creates wells with a depth between ~ 600-900 nm. In the final step, we remove the photoresist, in a bath of 1165 Remover where the chip soaks for 12 hours at 110 °C. To further clean the chips of photoresist residue, we put the etched wells into an $O_2$ plasma at 300 Watts with 300 sccm of $O_2$ for 2 minutes. The microcavities are then placed into a thermal evaporator. Germanium, Chromium, and Titanium are evaporated over the etched well with a thickness of 300 Å and a rate of 0.5 Å/s. For the gold substrate, 180 Å-thick Titanium is used as an adhesive and 450 Å of gold is subsequently evaporated. A similar process is utilized for the other metals studied (Fig. S2).

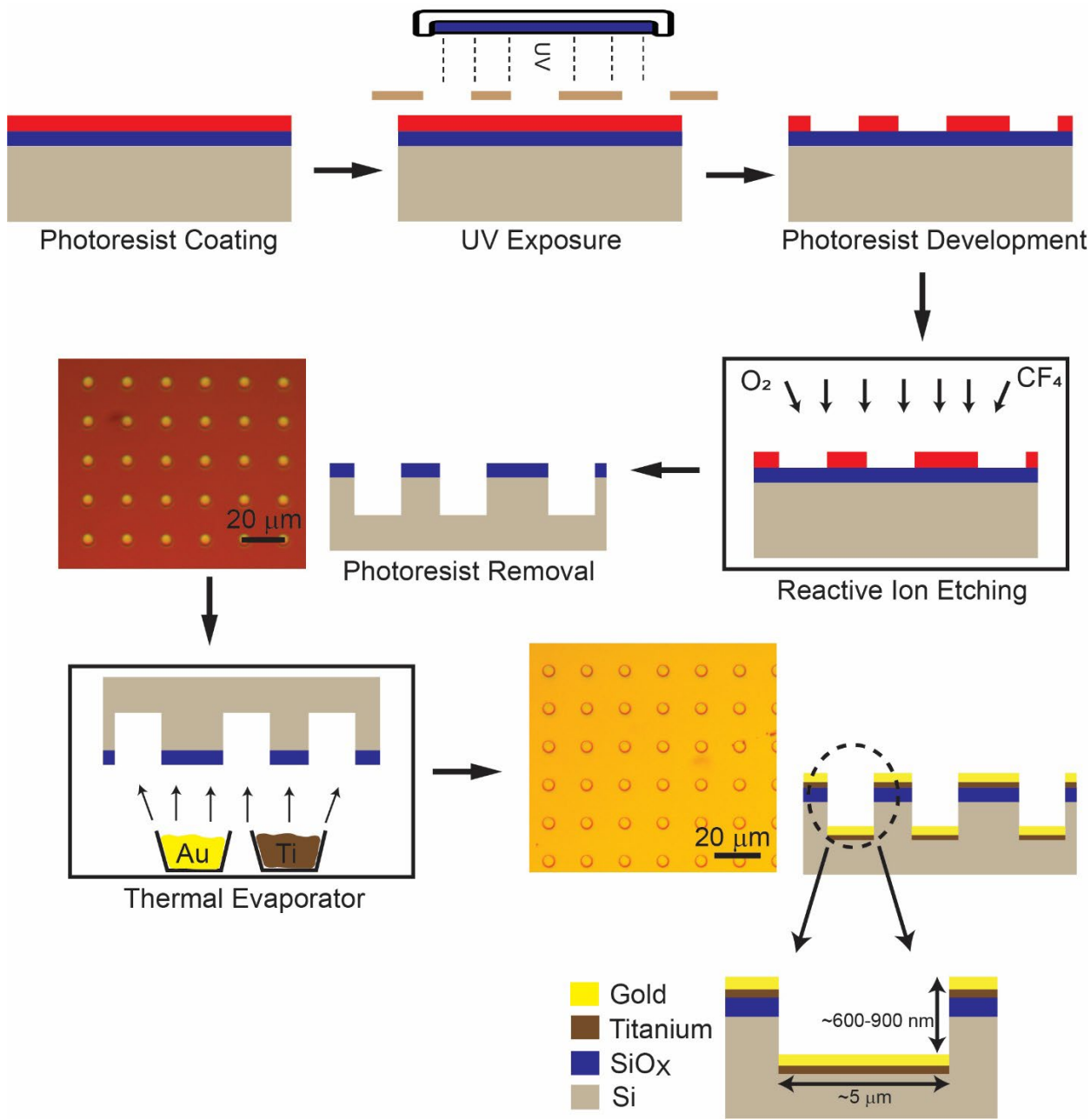

**Figure S2.** Microfabrication of the metal wells.

## 3. Graphite Substrate Fabrication

The preparation of the graphite wells starts with cleaning the $SiO_x$ surface with acetone, IPA, and DI water. With the 'Scotch-Tape' method, freshly exfoliated flakes on tape are pressed against the oxide surface. The tape is peeled from the surface very slowly (~ 1 mm/min). The substrate is then spin-coated with S1818 at 2500 rpm and kept on the hot plate at 115 °C for 1 minute. The spin-coated chips are exposed to UV for 20 seconds with 8 mW power. Again, we use the same patterned mask during UV exposure to form the circular wells. Using RIE, we etch the wells to ~ 600-900 nm deep by deploying the same etching parameters as described above for the metal devices. Devices are then placed into the Remover 1165 bath at 110 °C over 12 hours and then exposed to $O_2$ plasma to remove any remaining photoresist (Fig. S3).

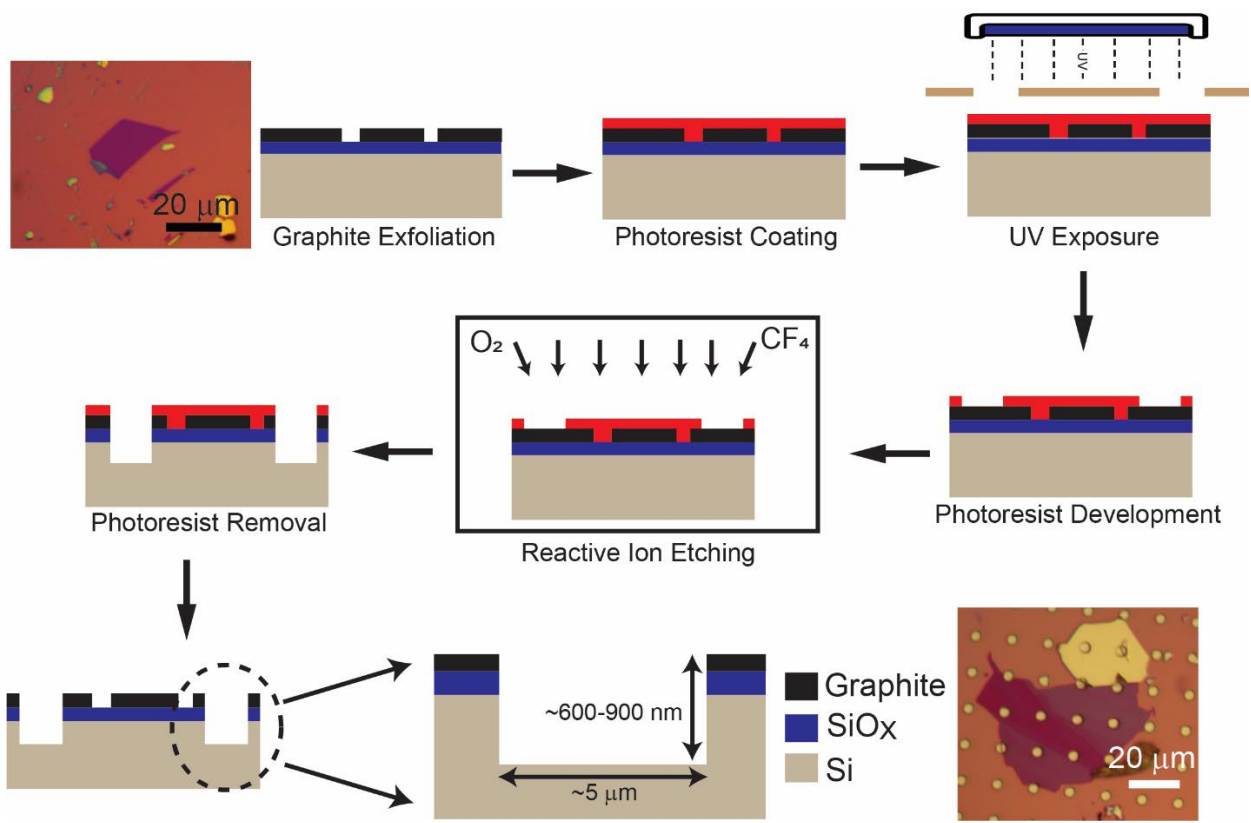

**Figure S3.** Microfabrication of the graphite wells.

## 4. Graphite Surface Treatment

Due to the $O_2$ plasma exposure for cleaning the graphite surface after photolithography, we performed Raman spectroscopy to determine if any surface chemistry changes took place. First, we carried out Raman spectroscopy over the freshly exfoliated flake (Fig. S4a) and again after photolithography and exposure to $O_2$ plasma during the final cleaning step (Fig. S4b). As can be seen in Figure S4b, a small D peak is introduced after $O_2$ plasma exposure indicative of the formation of defects[6] or graphite oxide formation[7] in the graphite lattice. However, the small size of the D peak suggests that the surface is not heavily oxidized.

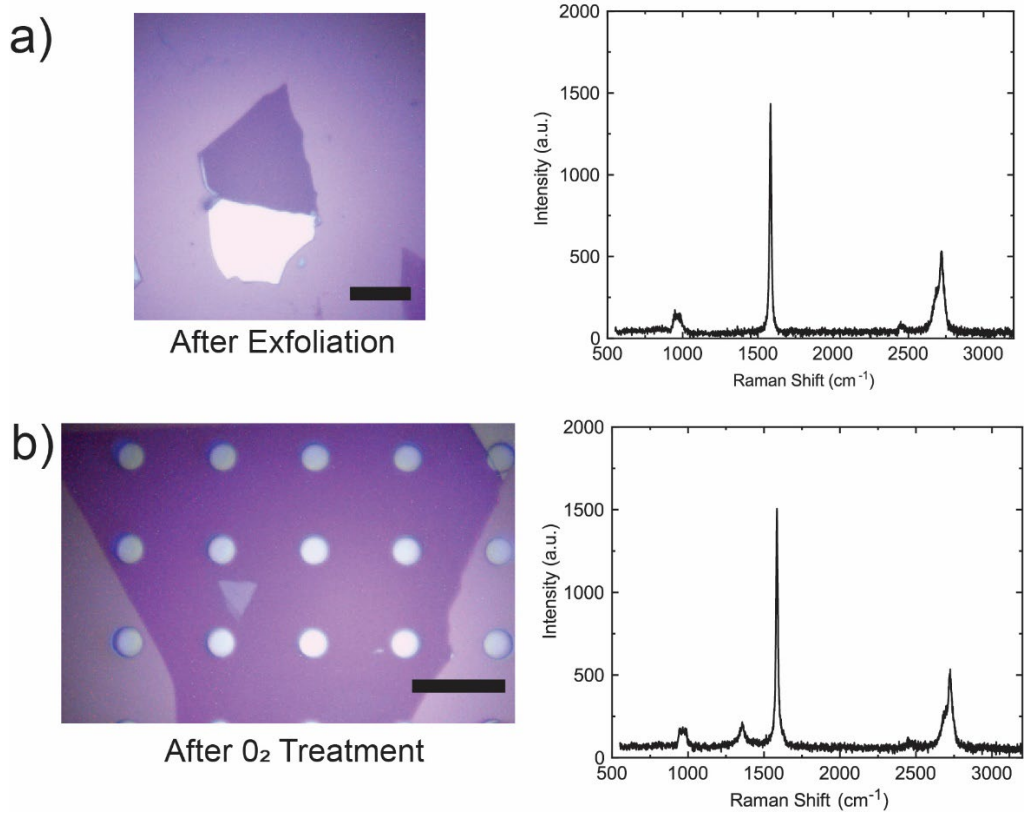

**Figure S4. a)** Optical image and Raman Spectrum of the freshly exfoliated graphite flake. Scale bar is 30 μm. **b)** Optical image and Raman spectroscopy of the etched FLG wells which exposed to $O_2$ plasma. The peak at 1400 cm$^{-1}$ due to defects. Scale bars are 20 μm.

## 5. MoS$_2$ Transfer Procedure

The transfer of MoS$_2$ begins by spin coating the CVD-grown MoS$_2$ flakes with PMMA at 2500 rpm. We create a window on the thermal-release tape and it is stamped onto the PMMA-covered MoS$_2$ substrate. Utilizing the hydrophilic nature of the SiO$_x$/MoS$_2$ interface, we put this combination into water and let the water separate the MoS$_2$ flakes from the SiO$_x$ substrate. This leaves us with a MoS$_2$/PMMA/Thermal-Release-Tape (MPT) combination.

Following the preparation of the MPT combination, metal and graphite substrates are both placed on a hot plate at 85 °C, and MPT is put onto the target substrate. For the graphite well, the transfer is carried out under the optical microscope with a custom-made apparatus that is used to keep the MPT still while approaching to surface. The Thermal-Release-Tape is peeled off from the surface easily with heat and the MoS$_2$/PMMA then sticks to the substrate. Before the annealing process, we put this device into a desiccator overnight (< 12 hr) to avoid any damage that trapped air inside the cavity may cause. The annealing process is carried out at 350 °C for 7 hours and under 20 sccm of Argon flow to help remove the PMMA from the surface (Fig. S5a). The transfer results are shown in the Fig. S5b (graphite) and Fig. S5c (gold).

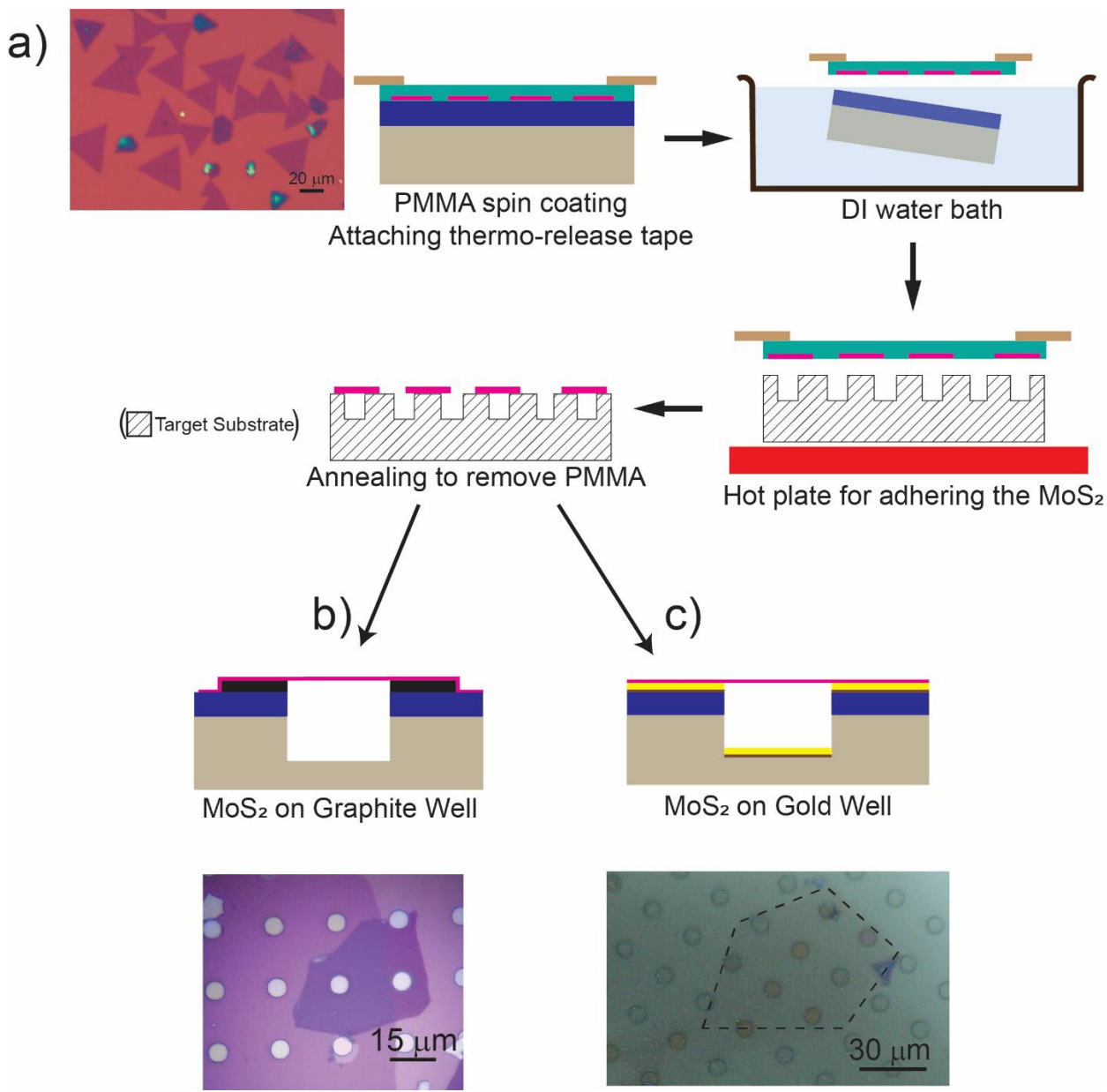

**Figure S5. a)** Schematic of monolayer MoS$_2$ flakes transfer over the target substrate. **b-c)** Optical image and schematic of the final result of the MoS$_2$ transfer over the graphite well and gold wells respectively. The dashed line in (c) shows the boundary of the MoS$_2$ flake.

# 6. Deriving the Expression for Pressure Difference and Deflection

Following the Hencky's solution for the deformation of a pressurized clamped axisymmetric membrane[8,9], we obtain the deflection profile as an infinite summation of even powers of normalized radius;

$$z(r) = a \left(\frac{\Delta p \, a}{Et}\right)^{1/3} \sum_{n=0}^{\infty} A_{2n} \left(1 - \left(\frac{r}{a}\right)^{2n+2}\right)$$

S6.1

with $A_0 = 1/B_0$, $A_2 = 1/2B_0^4$, $A_4 = 5/9B_0^7$, $A_6 = 55/72B_0^{10}$, $A_8 = 7/6B_0^{13}$, and so on. $E$ is the bulk Young's modulus, $t$ is the thickness of the membrane, and $\Delta p$ is the pressure difference. The maximum height of the blister at the center can be found by $\delta \equiv z(r = 0)$ which yields:

$$\delta = a \left(\frac{\Delta p \, a}{Et}\right)^{1/3} \sum_{n=0}^{\infty} A_{2n}$$

S6.2

By setting $K(v) = 1/(\sum_{n=0}^{\infty} A_{2n})^3$ and rearranging Eqn. S6.2, the final expression results in Eqn.1 of the main text.

One assumption of the constant-N blister test is the expansion of the membrane occurs isothermally, so we can model the change in the volume by ideal gas law which gives $p_0 V_0$ $= p_{int} (V_0 + V_b)$, where $V_0$ is the volume of the microcavity and $\Delta p = p_{int} - p_{ext}$. From Hencky's solution, the volume under the bulge is determined by the formula $V_b = C(v)\pi a^2 \delta$ where $C(v = 0.29) = 0.552$ since $v = 0.29$ [9,10] for MoS$_2$.

Below, we tabulate the constants $C, K$ for a range of MoS$_2$ Poisson's ratio values commonly found in the literature (Table S1) and the corresponding calculated 2D Young's modulus and work of separation using these values. As can be seen, it does not influence the work of separation and has only a minor effect on the calculated elastic modulus.

| | Poisson's Ratio (*v*) | K(*v*) | C(*v*) | E$_{2D}$ (N/m) | Γ$_{sep}$ (J/m²) |
|---|---|---|---|---|---|
| **Device ID: R13** | 0.29 | 3.54 | 0.522 | 218.8 | 0.32 |
| | 0.25[11] | 3.39 | 0.523 | 228.3 | 0.32 |

Table S1. Values for the constants C and K obtained with different Poisson's ratios and the E$_{2D}$ and Γ$_{sep}$ calculated using the corresponding constants.

# 7. Determining the Young's Modulus of MoS₂ over Graphite Devices

We examined 44 devices of MoS$_2$ over graphite substrates. In Fig. S6a, we plot $K(v)\,\delta^3/a^4$ vs $\Delta p$ for 8 representative devices. In Fig. S6b, we show the full data set of the calculated $E_{2D}$ values which were used to determine the work of separation (Fig. 4a in main text). Figure S6c shows an optical image of the graphite devices pinned to their initial radii, while the nearby membranes over the SiO$_x$ wells show delamination at the same input pressure. Figure S6d-e are AFM images of those same devices showing that the MoS$_2$ over the graphite wells remains adhered while those on the SiO$_x$ are delaminated for the same input pressure confirming the lower work of separation for MoS$_2$/SiO$_x$.

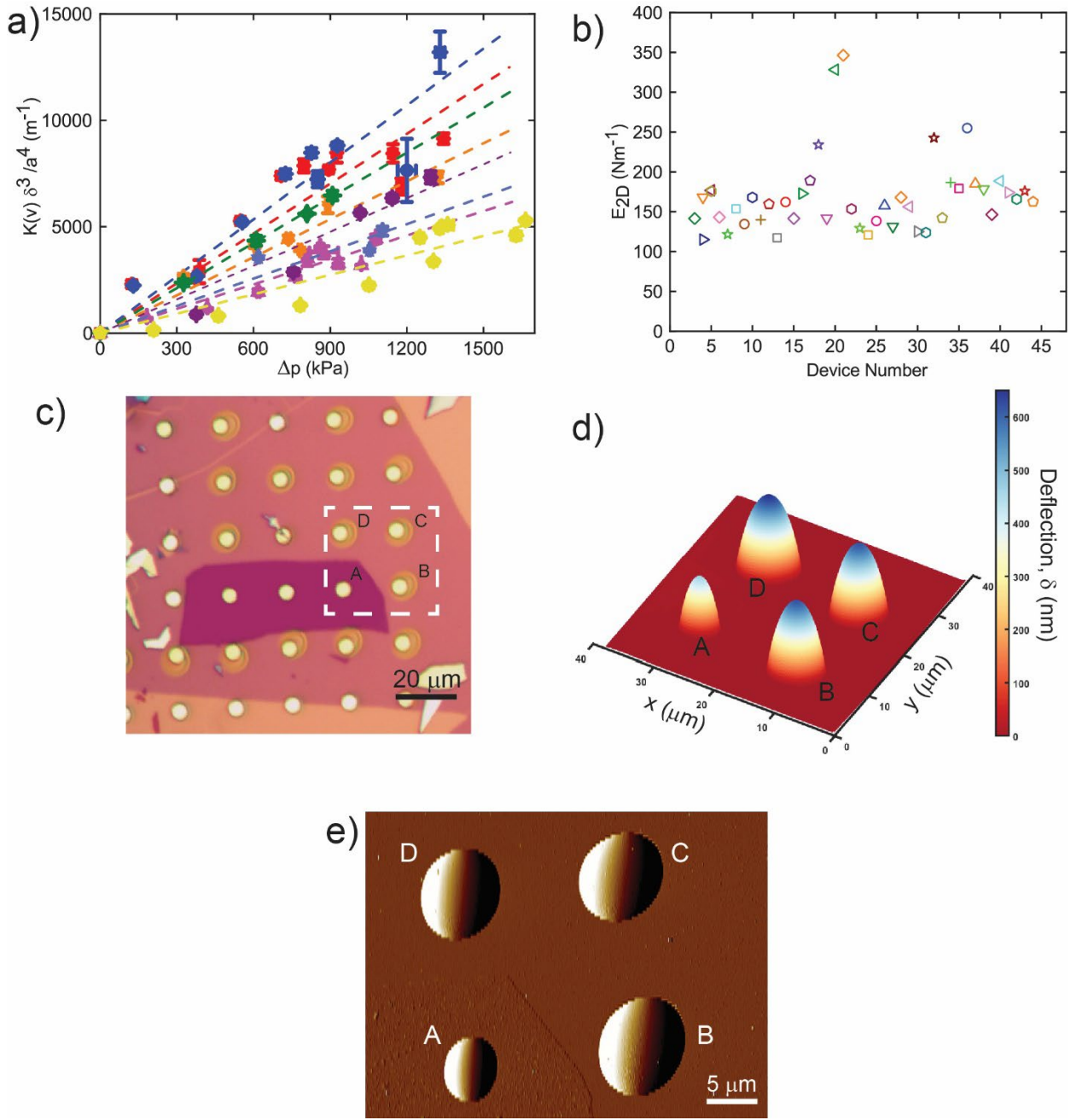

**Figure S6. a)** $K(v)\delta^3/a^4$ vs $\Delta p$ for CVD-grown MoS$_2$ membranes. Dashed lines are the linear fits used to determine E$_{2D}$ of each device. **b)** E$_{2D}$ for each device. **c)** Optical image of MoS$_2$ over the graphite and SiO$_x$ substrates (Device A is on the graphite and Devices B, C, and D are on the SiO$_x$). **d)** 3D AFM image of the devices labeled at (c). **e)** AFM Amplitude image of the devices labeled in (c).

# 8. Using Photoluminescence to Verify the Clamping Condition Assumed in Hencky's Model

MoS$_2$ shows changes in the optical band gap of ~ 100 meV/ % for biaxial strain[1,12]. Our membranes are subjected to biaxial strain near the center via a pressure difference $\Delta p = p_{int} - p_{ext}$. To validate the assumption in the Hencky model of a perfectly clamped circular membrane we used strain measurements determined by PL measurements and correlated them with strain measurements made by the AFM. A line scan of a series of PL measurements over the blister is shown in Fig. S7a. These are used to accurately determine the strain at the center of the membrane. In Figure S7b, we plot the PL response that corresponds to the points labeled with different colors indicating the different points (Fig. S7c, and d). In addition, we fit the Voigt function to each spectrum to follow the A exciton peak position[13]. We repeated the PL measurements for every new input pressure until delamination, and AFM scans accompany those measurements. We then correlated the AFM scans and PL measurements[8,9,14]. We use the stress-strain governing equations

$$\sigma_r - \nu\sigma_\theta = Et\varepsilon_r$$

S8.1

$$\sigma_\theta - \nu\sigma_r = Et\varepsilon_\theta$$

S8.2

where $\sigma_r$ and $\sigma_\theta$ are radial and tangential membrane stresses respectively and $\varepsilon_r$ and $\varepsilon_\theta$ are the radial and tangential strains. At the center of the membrane where $r = 0$, $\varepsilon_\theta = \varepsilon_r$ which is equivalent to perfect biaxial strain ($\varepsilon_b$). We can also write the equation for $\sigma_r$ as an infinite series of even powers of radius $r$,

$$\sigma_r = \left(\frac{Et\Delta p^2 a^2}{64}\right)^{1/3} \sum_{n=0}^{\infty} B_{2n} \left(\frac{r}{a}\right)^{2n}$$

S8.3

with $B_2 = -1/B_0^2$, $B_4 = -2/3B_0^5$, $B_6 = -13/18B_0^8$, $B_8 = -17/18B_0^{11}$, etc.

If we combine Eqn. S6.2 and S8.3 and Eqn. 1 from the main text, the final expression yields:

$$\varepsilon_b = \frac{B_0(\upsilon)(1-\upsilon)K(\upsilon)^{2/3}}{4} \left(\frac{\delta}{a}\right)^2$$

S8.4

where $B_o$ and $K$ depend on Poisson's ratio and $K(\upsilon) = 1/(\sum_0^{\infty} A_{2n})^3$. Using the Poisson's ratio for MoS$_2$ ($\upsilon = 0.29$), we obtain $B_o = 1.72$ and $K = 3.54$. We then substitute these values back into Eqn. S8.4 to obtain:

$$\varepsilon_b = 0.709 \left(\frac{\delta}{a}\right)^2$$

S8.5

From this expression, and knowing the maximum deflection and radius of the blister we calculate the biaxial strain at the center.

In Figure S7e, we plot the PL at the center of the blister vs. the strain calculated from the corresponding AFM scan for MoS$_2$ over the gold wells. The change in strain agrees with the theoretical change (~ 100 meV/%) validating the assumption of a perfectly clamped circular membrane assumed in the Hencky model. The laser is focused to a diffraction limited spot size of ~ 0.5 μm. The lateral spatial resolution is therefore ~ 0.5 μm. During the PL measurement, laser is kept focused over the scanned area.

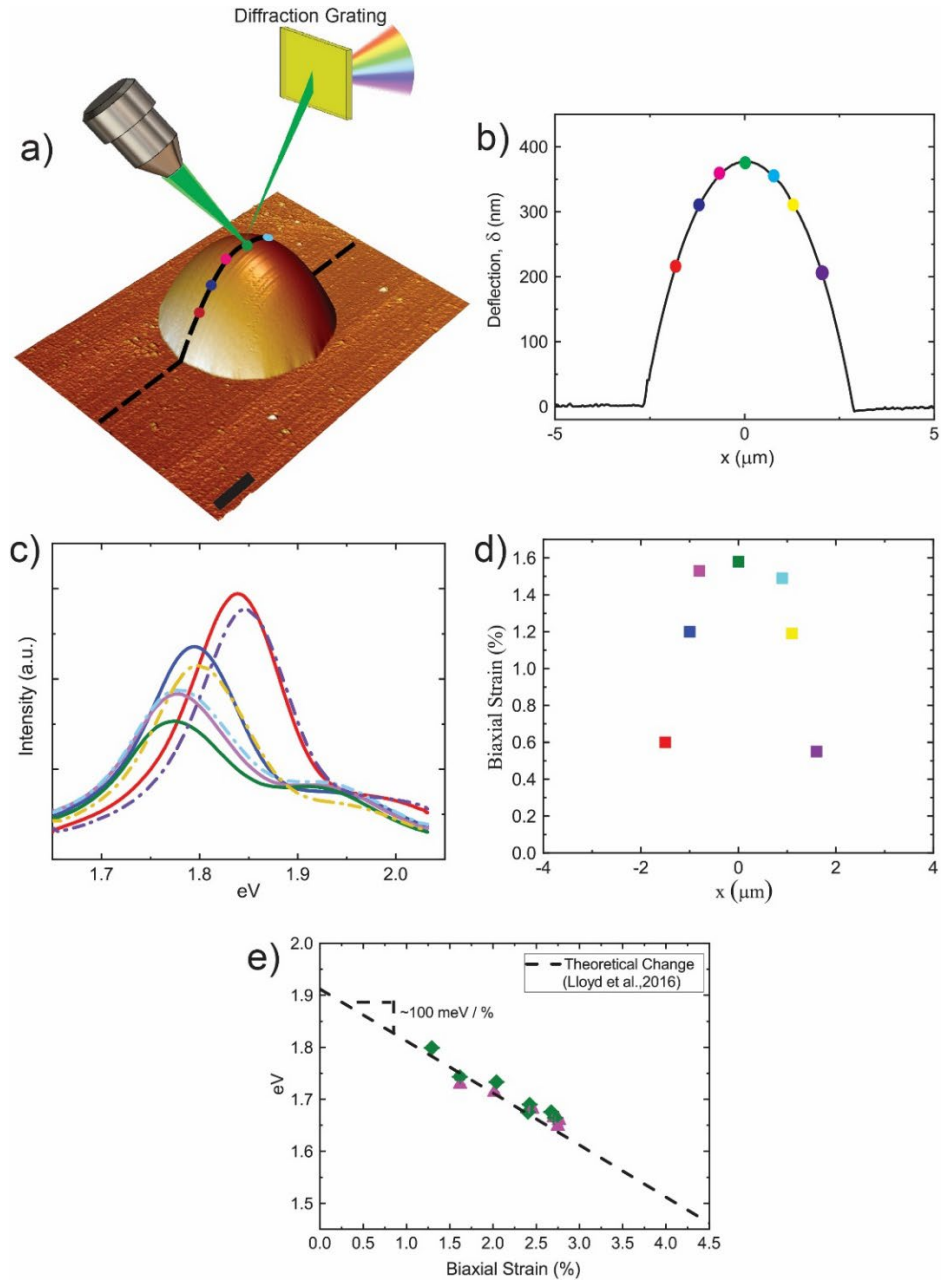

**Figure S7. a)** AFM image of the blister (height channel). The dashed line passes through the center of the blister where we conduct PL measurement over the points which are located along both sides of the blister. The scale bar is 2 μm. **b)** Positions of the PL measurement point on the cross-section of the device in (a). **c)** PL response of the data points labeled in (b) which we use the peak points in strain calculation. Solid lines correspond to points from the left edge of the blister to the center of the blister and dashed-dotted lines from the center to the right edge blister. **d)** Corresponding biaxial strain of the points in (c) by using their PL measurements. **e)** Calculation of the strain using the AFM results of each pressure change and matching them with corresponding PL data of $MoS_2$ on the gold wells. The dashed line is the theoretical change of the A exciton with respect to strain[1].

## 9. Strain Energy Derivation

In the paper, we closely follow the Wan et al. study[14,15] to calculate the contribution of the strain energy to our free energy model. First, we utilize equations described in the Hencky's solution.

Aside from the governing equations mentioned before (Eqn. S6.1, S6.2, S8.1, and S8.2), we also use the following equations:

$$\sigma_\theta = \frac{d}{dr}(r\sigma_r) \qquad \text{S9.1}$$

$$\varepsilon_\theta = \frac{u}{r} \qquad \text{S9.2}$$

where $u$ is radial displacement and $r$ is radial polar coordinate. Wan et al. showed that the strain energy in the membrane is equal to the work done by the external load, i.e, gas pressure while keeping the blister radius $a$ fixed.

$$F_{mem} = \int \Delta p(\delta) \, dV(\delta) \big|_a = \frac{\Delta p \, V_b}{4} \qquad \text{S9.3}$$

Assuming $\xi = \frac{r}{a}$, we obtain the volume of the bulge as,

$$V_b = \int_0^a z(r) 2\pi r \, dr = a^2 \left(\frac{\Delta p \, a^4}{Et}\right)^{\frac{1}{3}} \int_0^1 z(\xi) \, 2\pi\xi \, d\xi \qquad \text{S9.4}$$

Using S6.1, we get:

$$\text{S9.5}$$

$$F_{mem} = \frac{\Delta p\, a^2}{16}\left(\frac{\Delta p\, a^4}{Et}\right)^{1/3}\left(\frac{\pi}{8B_0} + \frac{\pi}{12B_0^4} + \frac{5\pi}{48B_0^7} + \frac{11\pi}{72B_0^{10}} + \frac{35\pi}{144B_0^{13}} + \frac{205\pi}{504B_0^{16}}\right.$$
$$+ \frac{17051\pi}{24192B_0^{19}} + \frac{2864485\pi}{2286144B_0^{22}} + \frac{20772653\pi}{9144576B_0^{25}} + \frac{135239915\pi}{32332608B_0^{28}}$$
$$\left.+ \frac{42367613873\pi}{5431878144B_0^{31}} + \cdots\right)$$

It can be shown that Eqn. S9.5 is equal to the conventional expression for membrane strain energy[16] in Eqn. S9.6.

$$F_{mem} = \int_0^a \left(\frac{1}{2}\sigma_r \varepsilon_r + \frac{1}{2}\sigma_\theta \varepsilon_\theta\right) 2\pi r\, dr \qquad \text{S9.6}$$

We obtain $\varepsilon_r$ and $\varepsilon_\theta$ by manipulating S8.1 and S8.2, respectively. After defining non-dimensional tangential stress component $\left(\sigma_\theta = \frac{d}{dr}(r\sigma_r) \Rightarrow \sigma_\theta = \frac{d}{d\xi}(\xi\sigma_r)\right)$, if we plug these back into Eqn. S9.6, we get:

$$F_{mem} = \pi a^2 \int_0^1 \left(\sigma_r^2 + \sigma_\theta^2 - 2\nu\sigma_r\sigma_\theta\right)\xi d\xi$$

$$= \pi a^2 \int_0^1 \left(2(1-\nu)\sigma_r^2 + 2(1-\nu)\xi\frac{d\sigma_r}{d\xi} + \xi^2\left(\frac{d\sigma_r}{d\xi}\right)^2\right)\xi d\xi \qquad \text{S9.7}$$

$$= \pi a^2 \int_0^1 \left(\frac{d}{d\xi}\left((1-\nu)\xi^2\sigma_r^2\right) + \xi^3\left(\frac{d\sigma_r}{d\xi}\right)^2\right)d\xi$$

The clamped boundary condition gives us the following relation:

$$u(\xi = 1) = 0 \Rightarrow \sigma_\theta(\xi = 1) - \nu\sigma_r(\xi = 1) = 0 \Rightarrow$$

$$(1-\nu)\sigma_r(\xi = 1)^2 = -\xi\, \sigma_r(\xi = 1)\left(\frac{d\sigma_r}{d\xi}(\xi = 1)\right) \qquad \text{S9.8}$$

After integrating the first term in the integrand, the final formula yields:

$$F_{mem} = \pi a^2 \left(\int_0^1 \xi^3 \left(\frac{d\sigma_r}{d\xi}\right)^2 d\xi\right) - \pi\left(\sigma_r(\xi = 1)\frac{d\sigma_r(\xi = 1)}{d\xi}\right) \qquad \text{S9.9}$$

Computing the Eqn. S9.9 using the series expressions for $\sigma_r$ (Eqn. S8.3) gives us:

$$\begin{aligned}F_{mem} = \frac{\Delta p\, a^2}{16}\left(\frac{\Delta p\, a^4}{Et}\right)^{\frac{1}{3}} &\left(\frac{\pi}{8B_0} + \frac{\pi}{12B_0^4} + \frac{5\pi}{48B_0^7} + \frac{11\pi}{72B_0^{10}} + \frac{35\pi}{144B_0^{13}} + \frac{205\pi}{504B_0^{16}} + \frac{17051\pi}{24192B_0^{19}}\right.\\
&+ \frac{2864485\pi}{2286144B_0^{22}} + \frac{20772653\pi}{9144576B_0^{25}} + \frac{135239915\pi}{32332608B_0^{28}} + \frac{42367613873\pi}{5431878144B_0^{31}}\\
&\left.+ \frac{1120150157\pi}{76212576B_0^{34}} + \frac{35059666851235\pi}{1254763851264B_0^{37}} + \cdots\right)\end{aligned} \qquad \text{S9.10}$$

As can be seen, the series in Eqns. S9.5 and S9.10 are identical.

## 10. Relamination of MoS₂ over the Metal and Graphite Wells

In Figure S8, the complete deflation data for 9 devices of MoS$_2$ over gold substrates are shown. We obtain a mean value of $\Gamma_{adh} = 0.017 \pm 0.005$ J/m$^2$. Figure S8b is the optical image of the delaminated device, and Figure S8c-h are AFM images showing relamination. The adhesion hysteresis observed is reminiscent of AFM based nanoindentation studies. In these studies the hysteresis has been attributed to moisture[17,18], viscoelasticity[19], plasticity[20], and surface instabilities[21,22]. Similar mechanisms as these may be at work in our case as well. The measurements of the deflection and radius are performed along the cross-sections shown in Fig. S8c and only the mean values of these measurements are shown in Fig. S8a for clarity.

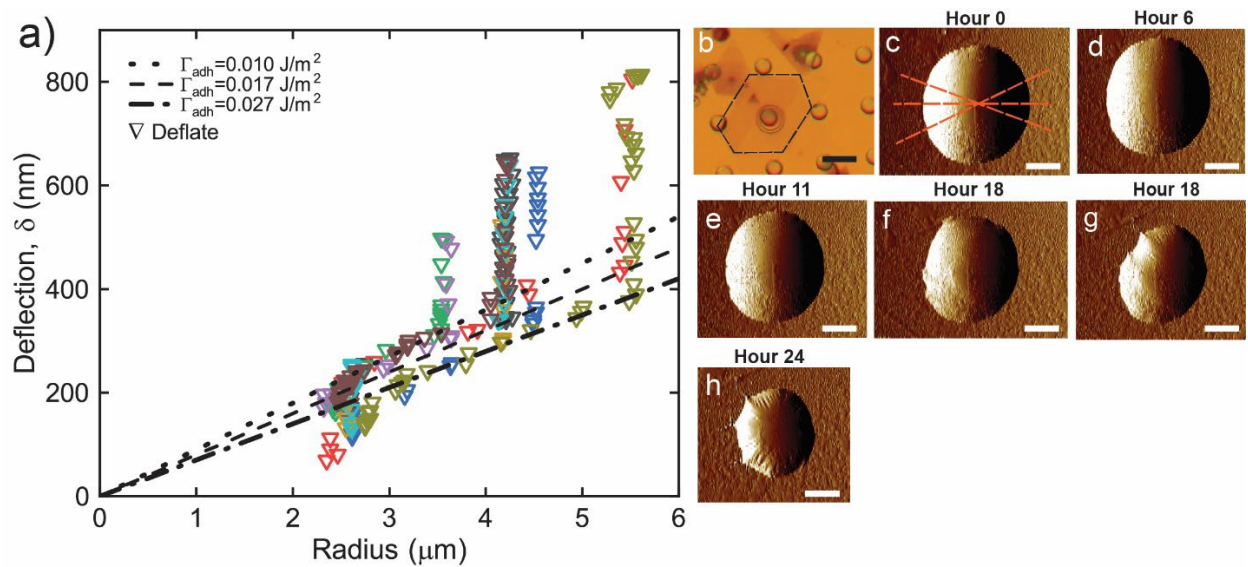

**Figure S8. a)** Complete data of deflation for 9 devices. Each color represents a different device. The dashed line is the mean value of all samples ($\Gamma_{adh} = 0.017 \pm 0.005$ J/m$^2$ (Dashed line). **b)** Optical image of the delaminated devices. The scale bar is 10 μm. **c-d)** AFM images(amplitude channel) of the device in (b). Elapsed times are put on top of each photo. The dashed lines in (c) are the cross-sections that are used for deflection and radius change. The scale bars are 3 μm.

In Figure S9, we plot the deflation data sets for the MoS$_2$ on the FLG substrates. We performed the measurements on 9 different devices and found $\Gamma_{adh}$ = 0.057 ± 0.008 J/m$^2$ (dashed line). Fig. S9b is the optical image of the devices and its deflation behavior is shown in Fig. S9c-k with each labeled with the elapsed time of the measurement.

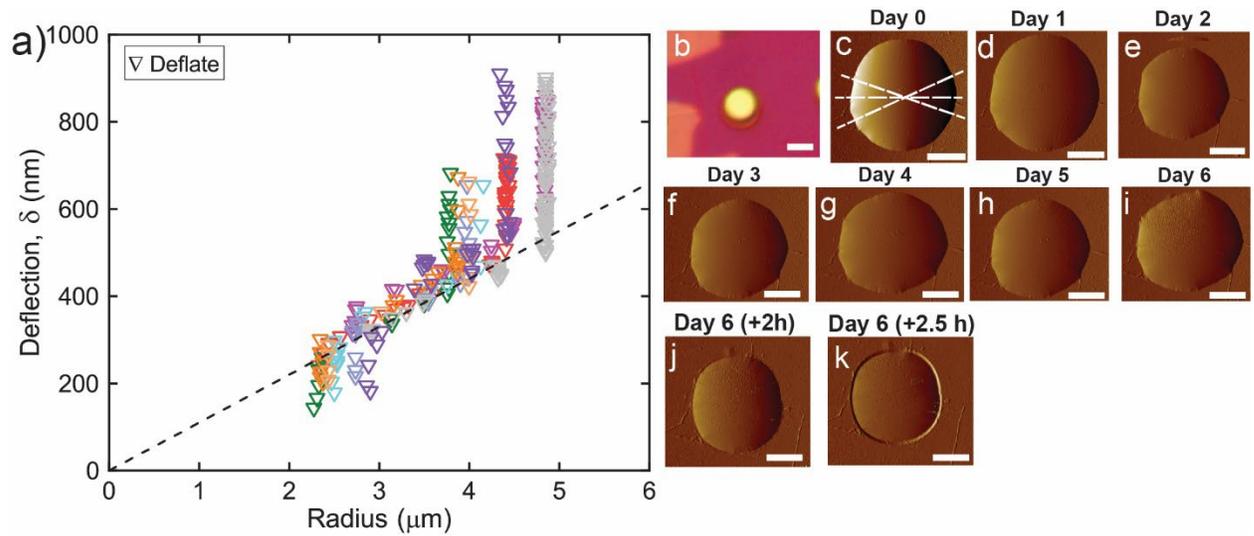

**Figure S9. a)** Complete deflation data for 9 graphite devices. Each color represents a different device. The dashed line is the mean value of all samples shown in Fig. 4d ($\Gamma_{rel}$ = 0.057 ± 0.008 J/m$^2$). **b)** Optical image of a delaminated device. The scale bar is 5 µm. **c-k)** AFM images (amplitude channel) of the device in (b). Elapsed times are included. The dashed lines in (c) are the cross-sections that are used for the measurement of deflection and radius. The scale bars are 3µm.

In Figure S10, we examine the MoS$_2$ over Si wells and in Fig. S10a we plot $K(v)\delta^3/a^4$ vs $\Delta p$. A linear fit to the data is used to determine $E_{2D}$ for each device. These calculated $E_{2D}$ values are used in the separation and adhesion energy calculations of these membrane from the Si wells. In Figure S10b, we show the values of $E_{2D}$ determined from the fit in Fig. S10a. Figure S10c is the deflation data for the MoS$_2$ on the Si substrates. We performed the measurements on 5 different devices. We found a work of adhesion $\Gamma_{adh} = 0.03 \pm 0.0018$ J/m$^2$ (dashed line).

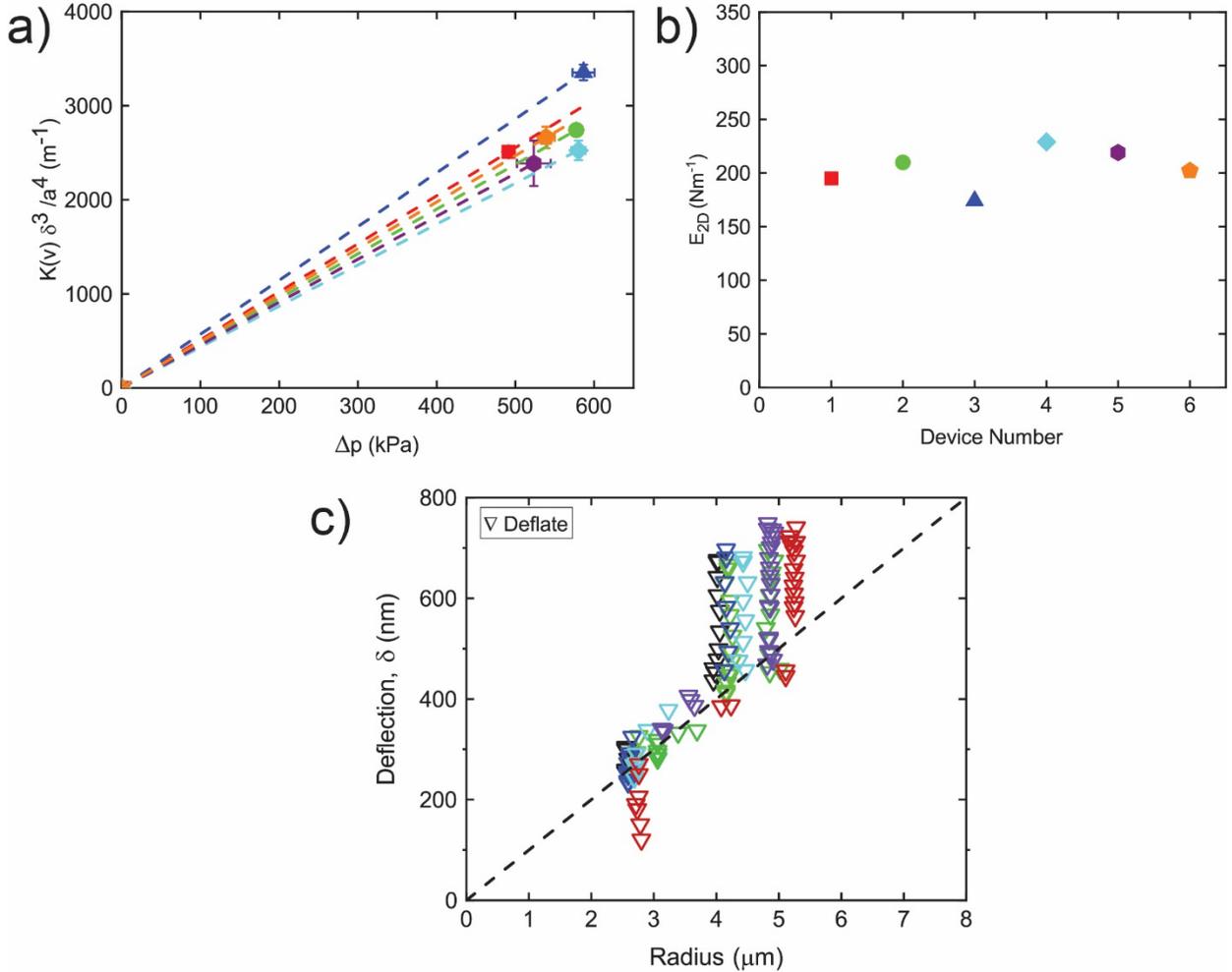

**Figure S10.** Data for Si substrates. **a)** $K(v)\delta^3/a^4$ vs $\Delta p$ for CVD-grown MoS$_2$ membranes. Dashed lines are the linear fits that are used to calculate $E_{2D}$ of each device. Symbols are color-coded. **b)** $E_{2D}$ for each device **c)** Maximum deflection, $\delta$, and radius during the deflation. The dashed line is the mean value of all samples.

In Figure S11 we examine the MoS$_2$ over SiO$_x$ wells and in Fig. S11a we plot $K(v)\delta^3/a^4$ vs $\Delta p$. A linear fit to the data is used to determine $E_{2D}$ for each device. In Figure S11b, we show the values of $E_{2D}$ determined from the fit in Fig. S11a. These calculated $E_{2D}$ values are used in the separation and adhesion energy calculations of these membrane from the SiO$_x$ wells.

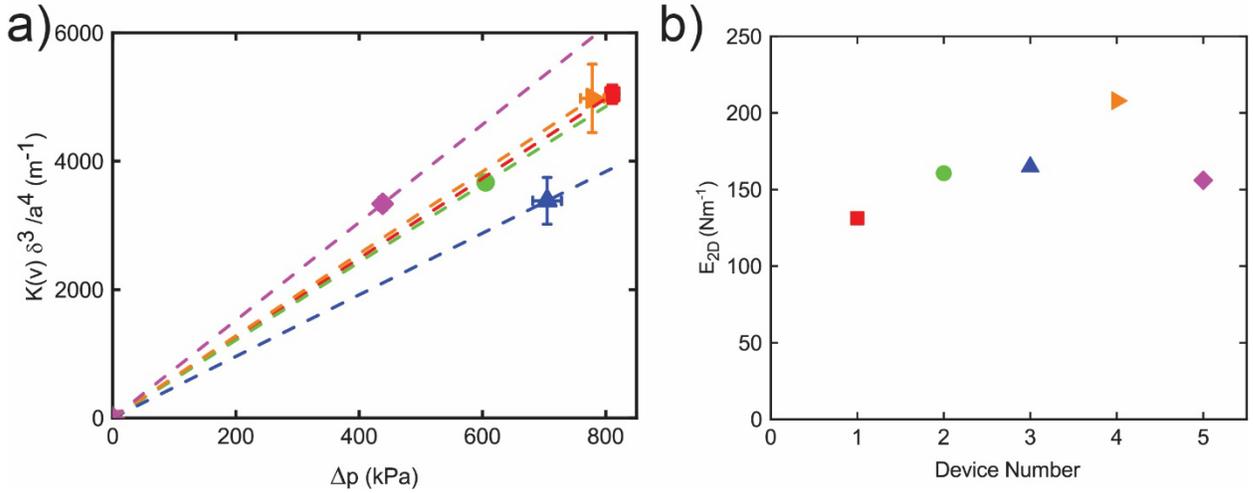

**Figure S11.** Data for SiO$_x$ substrates. **a)** $K(v)\delta^3/a^4$ vs $\Delta p$ for CVD-grown MoS$_2$ membranes. Dashed lines are the linear fits that are used to calculate E$_{2D}$ of each device. Symbols are color-coded. **b)** $E_{2D}$ for each device.

In Figures S12 through S14, we examine the MoS$_2$ over Titanium, Chromium, and Germanium wells. In each figure, we plot $K(v)\delta^3/a^4$ vs $\Delta p$. A linear fit to the data is used to determine $E_{2D}$ for each device. Later, we show the values of $E_{2D}$ determined from the fits which are used in the separation and adhesion energy calculations for each substrate. The average work of adhesion can be found in the caption of the figures for each material.

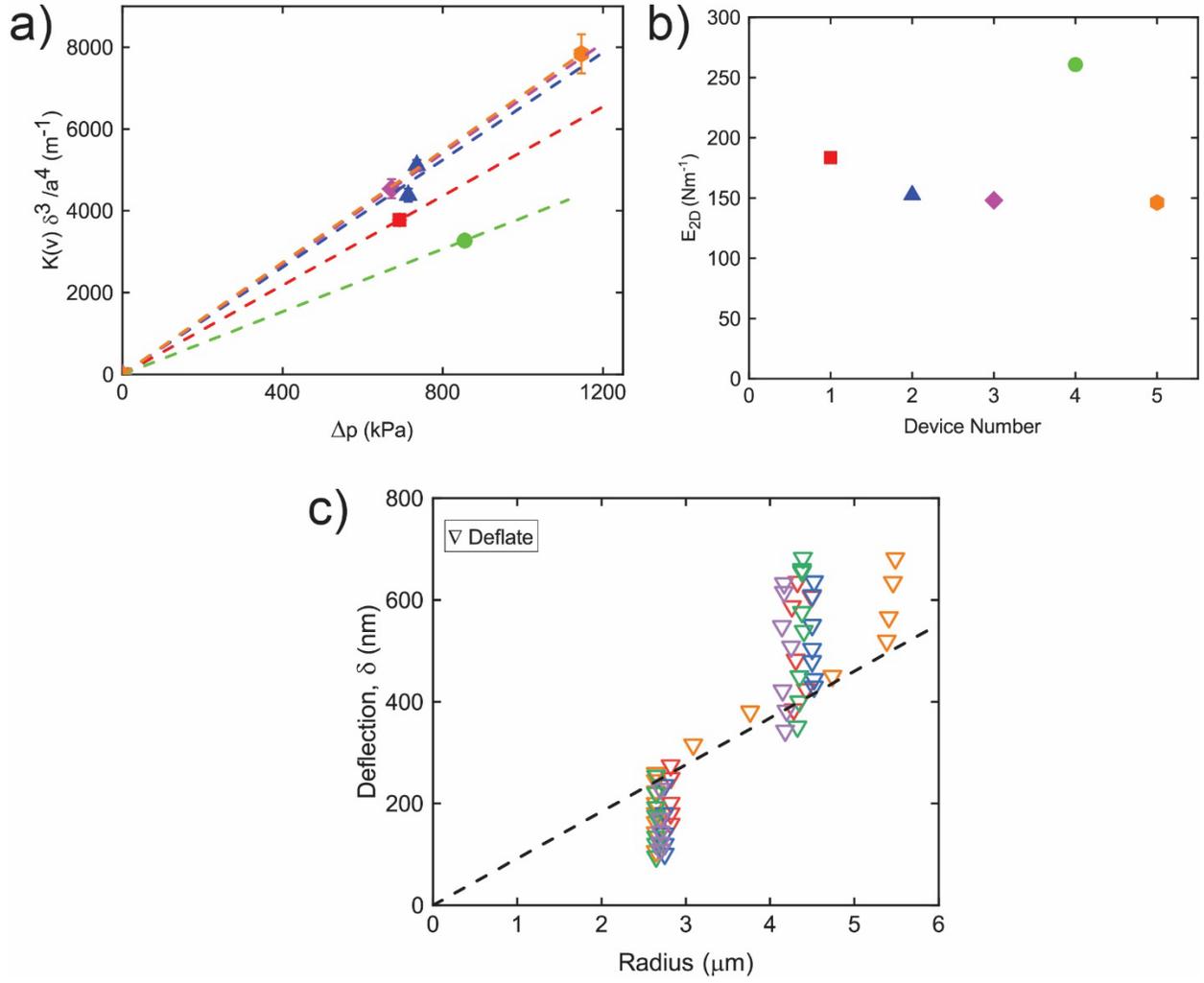

**Figure S12.** Data for Titanium substrates. **a)** $K(v)\delta^3/a^4$ vs $\Delta p$ for CVD-grown MoS$_2$ membranes. Dashed lines are the linear fits that are used to calculate E$_{2D}$ of each device. Symbols are color-coded. **b)** $E_{2D}$ for each device. **c)** $\delta$ vs radius during deflation. The dashed line is the mean value of all samples. $\Gamma_{adh} = 0.03 \pm 0.004$ J/m$^2$

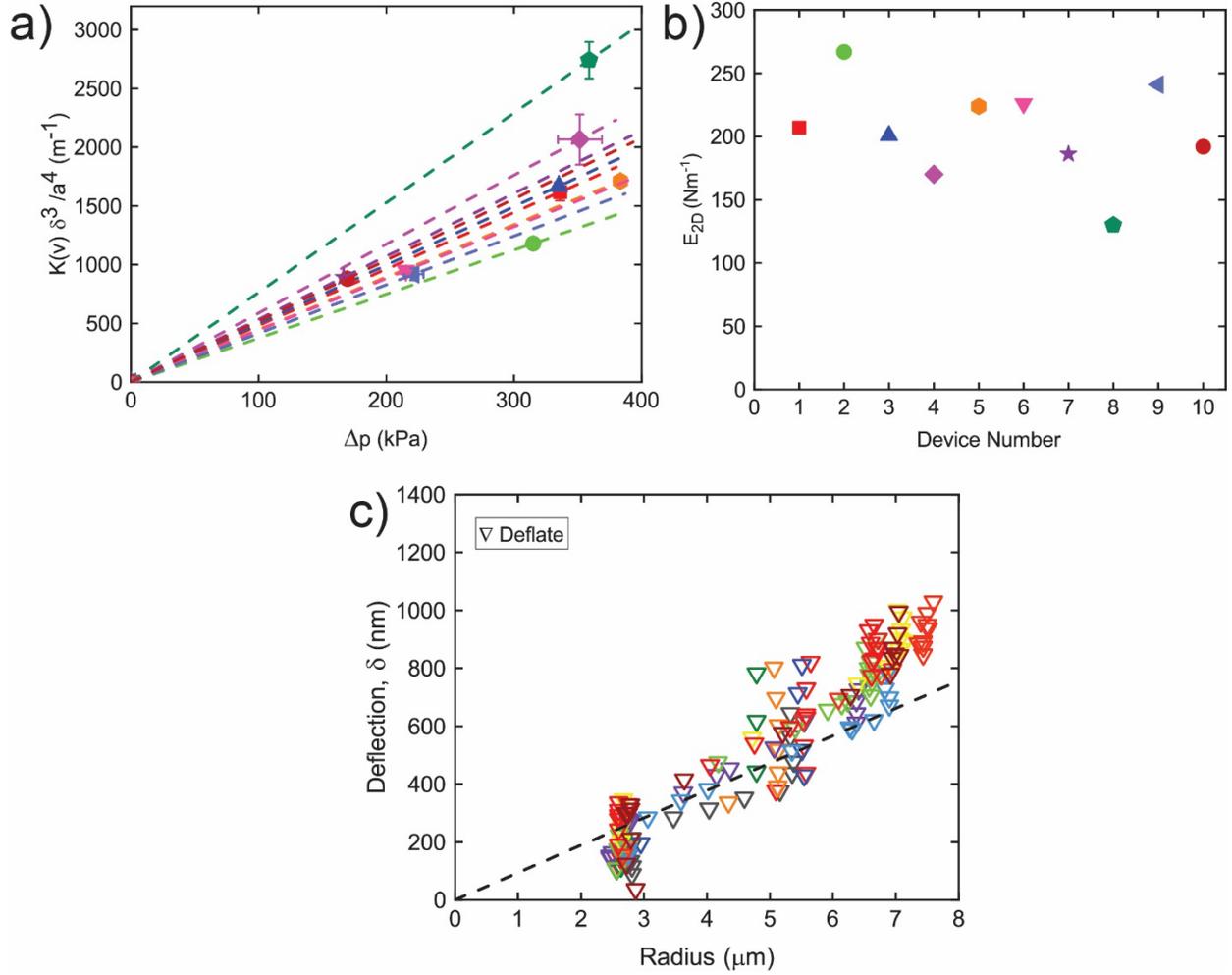

**Figure S13.** Data for Chromium substrates. **a)** $K(v)\delta^3/a^4$ vs $\Delta p$ for CVD-grown MoS$_2$ membranes. Dashed lines are the linear fits that are used to calculate E$_{2D}$ of each device. Symbols are color-coded. **b)** $E_{2D}$ for each device. **c)** $\delta$ and radius during deflation. The dashed line is the mean value of all samples. $\Gamma_{adh} = 0.036 \pm 0.018$ J/m$^2$.

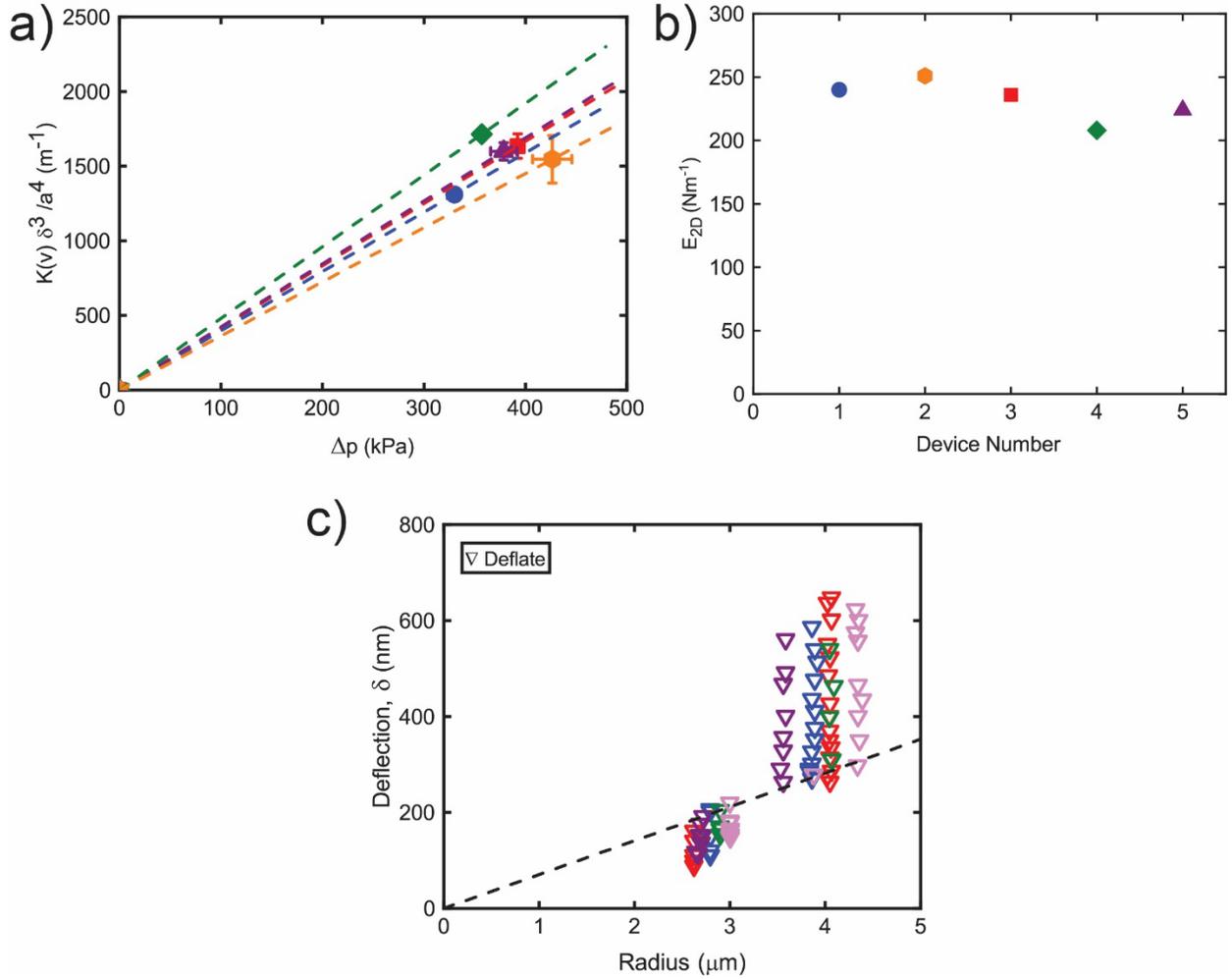

**Figure S14.** Data for Germanium substrates. **a)** $K(v)\delta^3/a^4$ against $\Delta p$ for CVD-grown MoS$_2$ membranes. Dashed lines are the linear fits that are used to calculate E$_{2D}$ of each device. Symbols are color-coded. **b)** $E_{2D}$ for each device. **c)** $\delta$ and radius during deflation. The dashed line is the mean value of all samples. $\Gamma_{adh} = 0.012 \pm 0.002$ J/m$^2$.

## 11. Roughness Measurements over the Substrates

To determine the influence of surface roughness on the work of separation , we measured the surface roughness of our substrates using an AFM. First, we performed the AFM scans after the fabrication of the substrate. Following the MoS$_2$ transfers over the wells, we performed another AFM scan over the MoS$_2$ covered areas. AFM scans are conducted in a 250 nm x 500 nm area in tapping mode. Then, we analyzed 100 nm x 100 nm sub-areas within the scans to determine the root-mean-square (rms) values of the roughness using the NanoScope Analysis 2.0 program. The same scan sizes, and parameters are used for all of the scans.

In the following graphs, we show the AFM height images and the representative line cut through the scan area to demonstrate the surface roughness for each substrate used in our experiment.

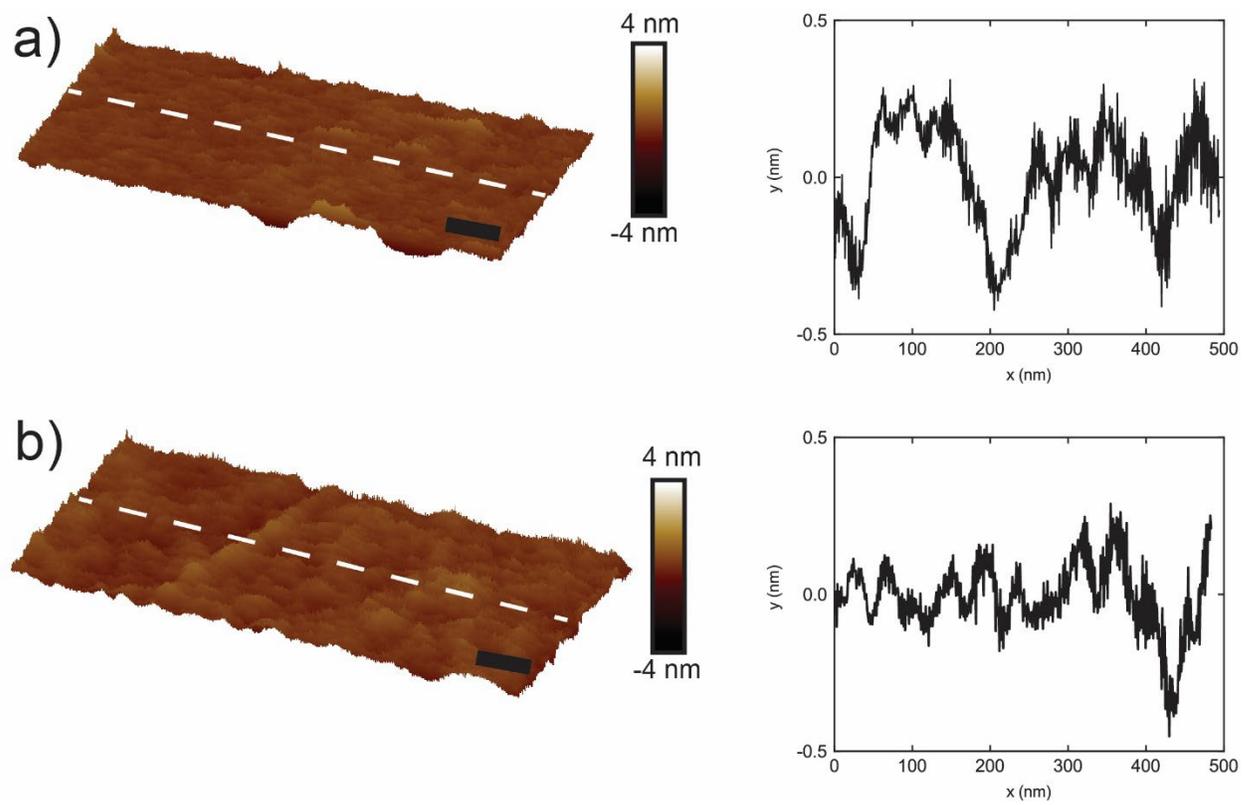

**Figure S15.** Si substrate. **a)** AFM scan over bare substrate fabrication of wells. **b)** Scan over MoS$_2$ covered area after annealing. Scale bars are 50 nm.

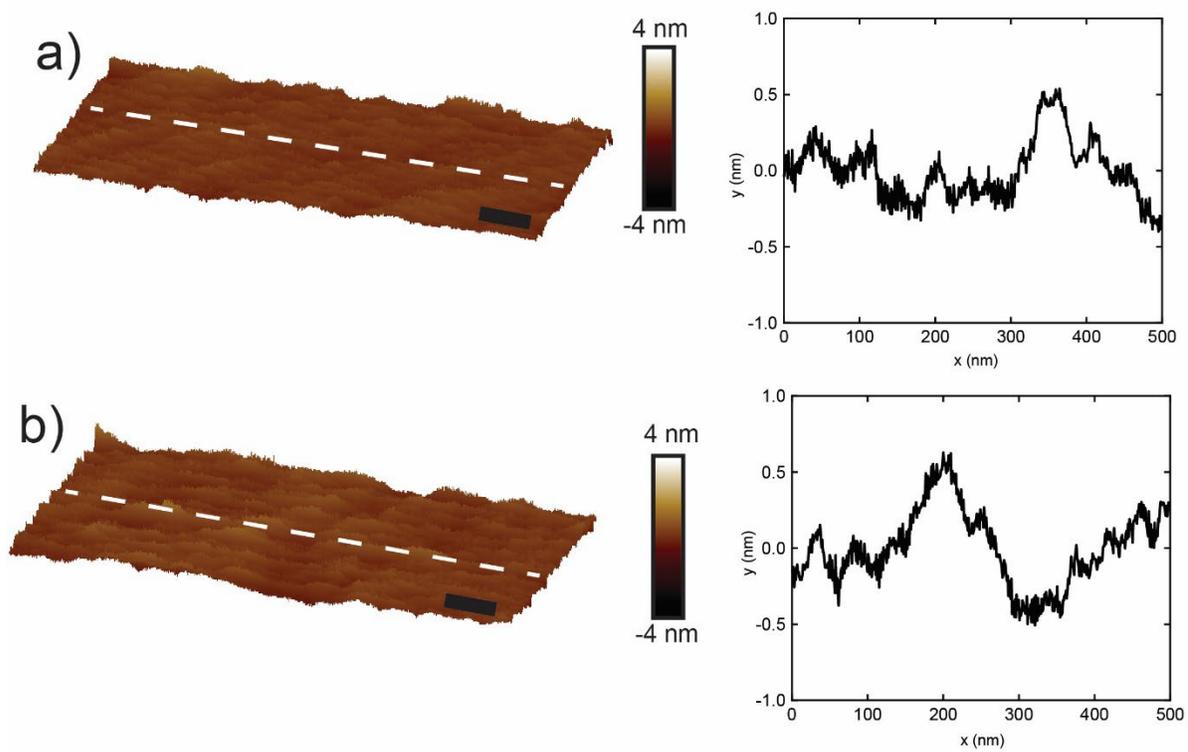

**Figure S16.** SiO$_x$ substrate. **a)** AFM scan over bare substrate after fabrication of wells. **b)** Scan over MoS$_2$ covered area after annealing. Scale bars are 50 nm.

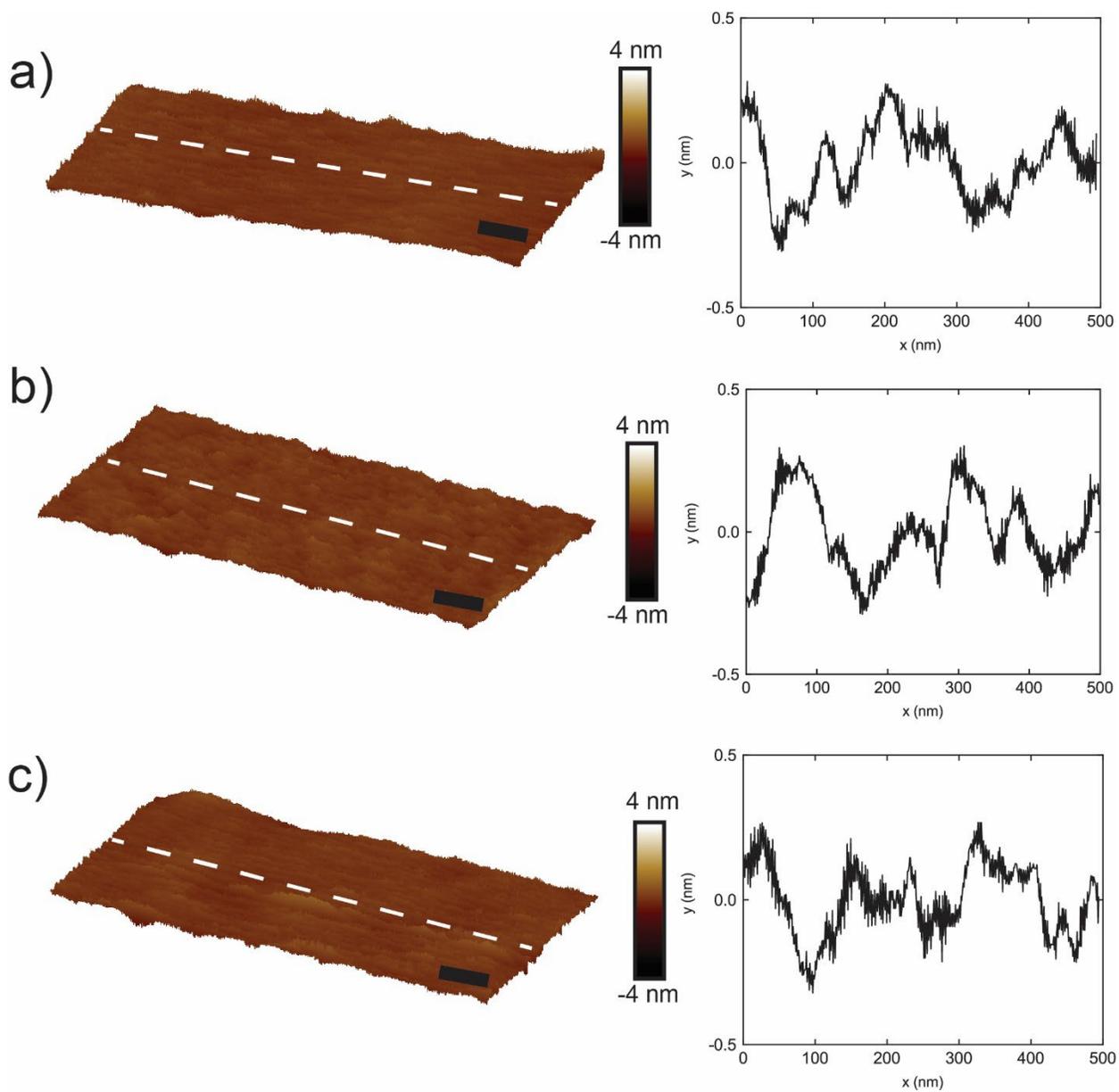

**Figure S17.** Graphite substrate. **a)** AFM scan after fabrication of wells. **b)** Scan over bare substrate after annealing. **c)** Scan over MoS$_2$ covered area after annealing. Scale bars are 50 nm.

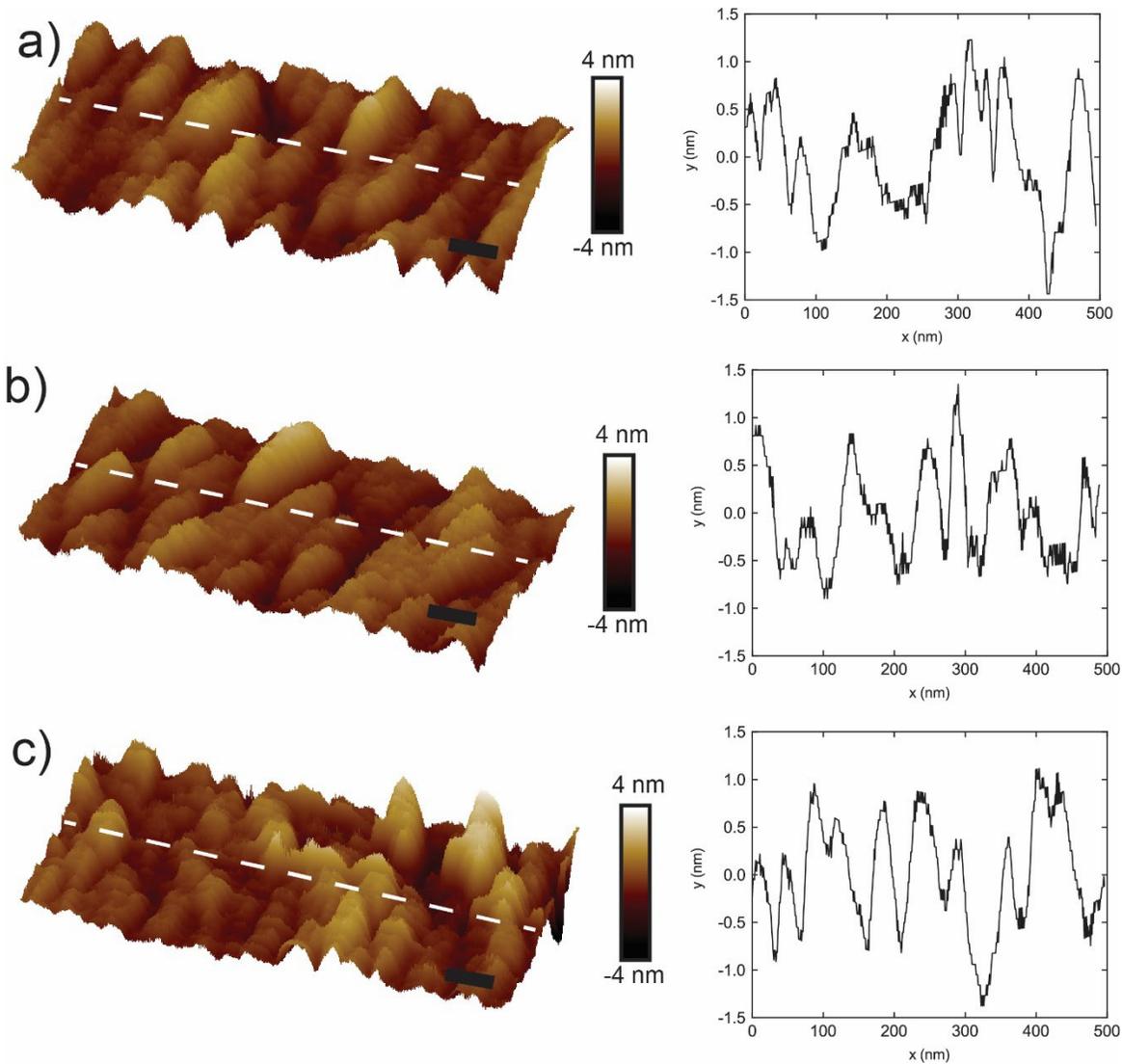

**Figure S18.** Gold substrate. **a)** AFM scan after fabrication of wells. **b)** Scan over bare substrate after annealing. **c)** Scan over MoS$_2$ covered area after annealing. Scale bars are 50 nm.

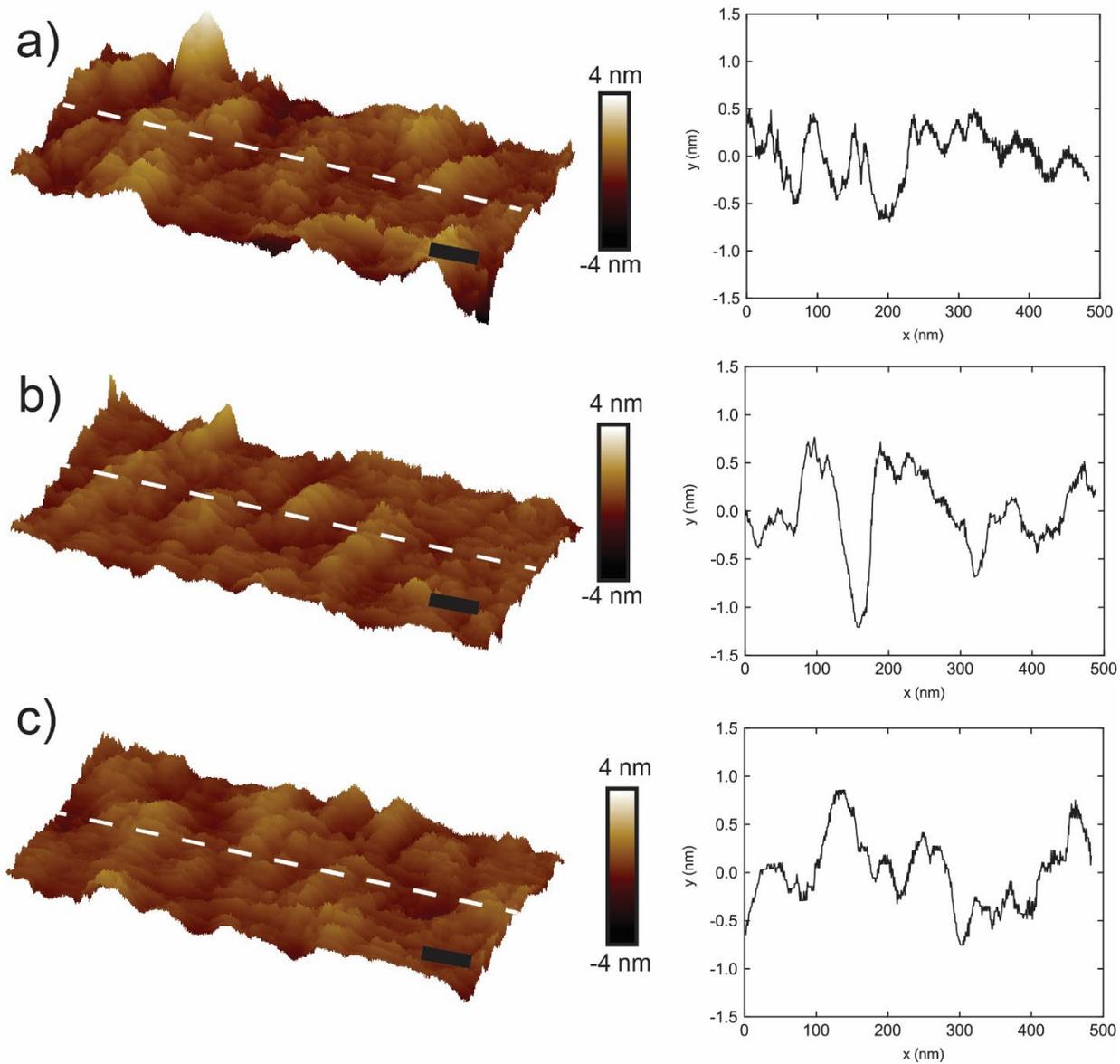

**Figure S19.** Titanium substrate. **a)** AFM scan after fabrication of wells **b)** Scan over bare substrate after annealing. **c)** Scan over MoS$_2$ covered area after annealing. Scale bars are 50 nm.

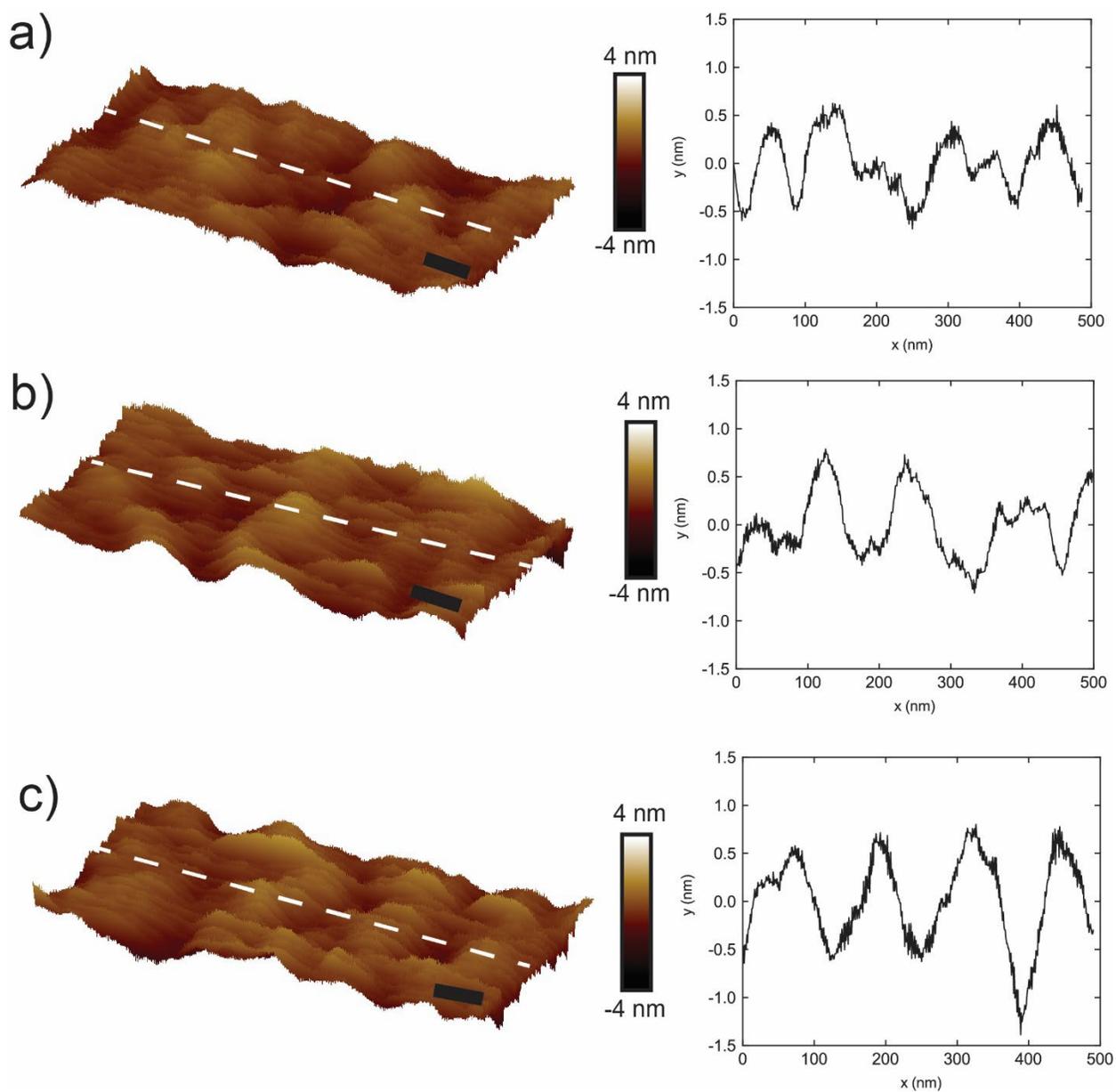

**Figure S20.** Chromium substrate. **a)** AFM scan after fabrication of wells **b)** Scan over bare substrate after annealing. **c)** Scan over MoS$_2$ covered area after annealing. Scale bars are 50 nm.

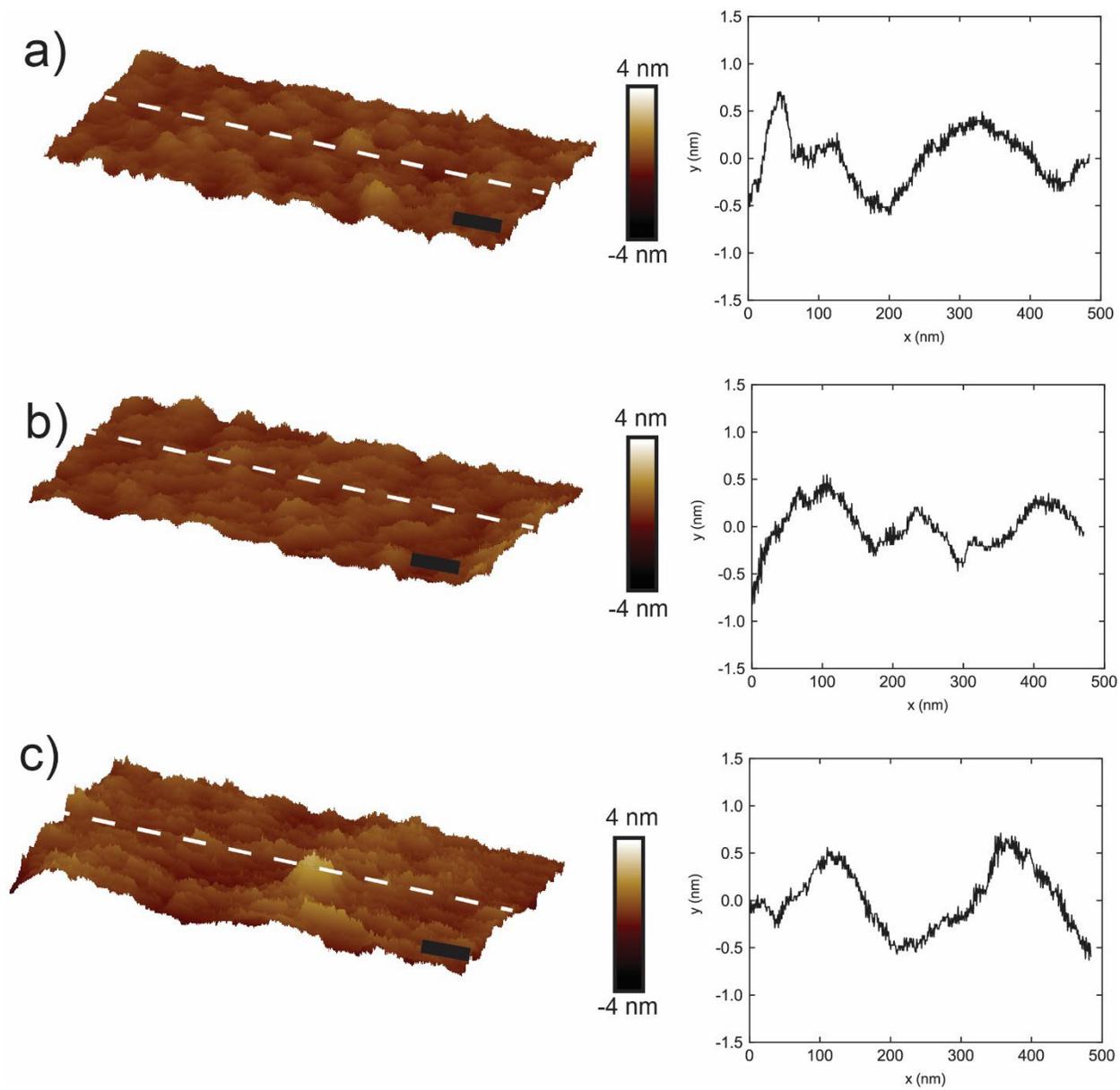

**Figure S21.** Germanium substrate. **a)** AFM scan after fabrication of wells **b)** Scan over bare substrate after annealing. **c)** Scan over MoS$_2$ covered area after annealing. Scale bars are 50 nm.

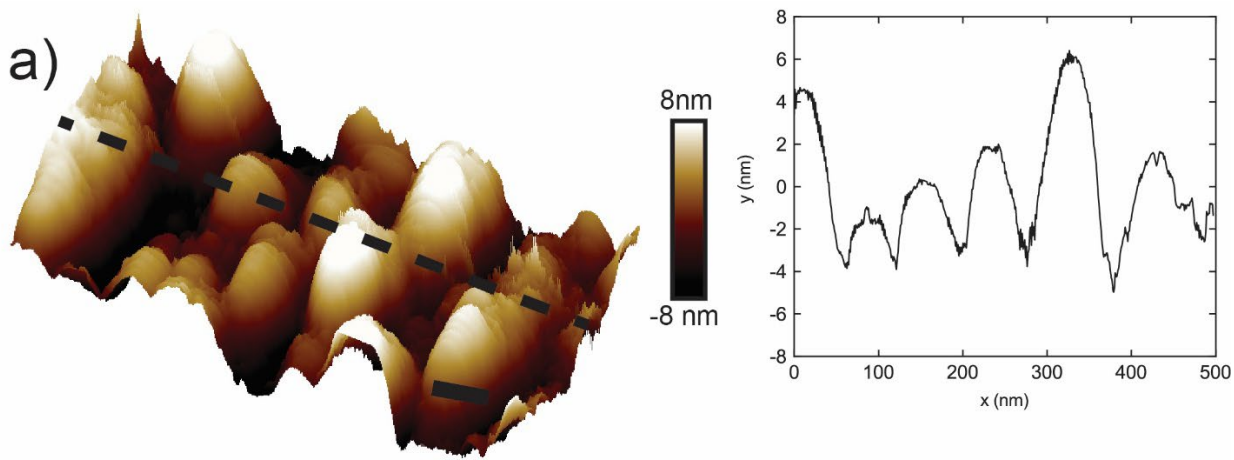

**Figure S22.** Aluminum substrate. **a)** AFM scan after fabrication of wells. Scale bar is 50 nm.

In Table S2, the results of the AFM scans over the substrates are tabulated.

| Substrate Material | Roughness Measurements (Averaged RMS Values) | | | Work of Separation (Jm$^{-2}$) | Work of Adhesion (Jm$^{-2}$) |
|---|---|---|---|---|---|
| | Over the Bare Substrate Before Annealing (nm) | Over the Bare Substrate After Annealing (nm) | Over the MoS$_2$ Flake (nm) | | |
| Si | 0.15±0.02 | N/A | 0.21±0.04 | 0.20±0.03 | 0.03±0.018 |
| SiO$_x$ | 0.18±0.01 | N/A | 0.20±0.02 | 0.22±0.01 | N/A |
| Graphite | 0.21±0.04 | 0.27±0.03 | 0.26±0.01 | 0.39±0.1 | 0.05±0.01 |
| Gold | 0.7 ±0.08 | 0.57±0.04 | 0.6 ±0.02 | 0.28±0.04 | 0.017±0.005 |
| Chromium | 0.48±0.02 | 0.45±0.03 | 0.5±0.04 | 0.11±0.05 | 0.036±0.018 |
| Germanium | 0.32±0.04 | 0.32±0.01 | 0.33±0.02 | 0.12±0.04 | 0.01±0.002 |
| Titanium | 0.4±0.08 | 0.35±0.02 | 0.36±0.1 | 0.14±0.03 | 0.033±0.004 |
| Aluminium | 3.8±0.8 | N/A | N/A | N/A | N/A |

**Table S2.** Summary of roughness measurements.

## 12. Transfer of MoS$_2$ Membranes over Various Substrates

Videos showing the transfer of the MoS$_2$ onto the Silicon Oxide (video 1), Chromium (video 2), Titanium (video 3), and Aluminum (video 4) surfaces. In video 4, we attempted to transfer MoS$_2$ over the aluminum coated surface but were unsuccessful. Aluminum had the highest surface roughness and MoS$_2$ was unable to be transferred.

**References**

(1) Lloyd, D.; Liu, X.; Christopher, J. W.; Cantley, L.; Wadehra, A.; Kim, B. L.; Goldberg, B. B.; Swan, A. K.; Bunch, J. S. Band Gap Engineering with Ultralarge Biaxial Strains in Suspended Monolayer MoS2. *Nano Lett.* **2016**, *16* (9), 5836–5841. https://doi.org/10.1021/acs.nanolett.6b02615.

(2) Wu, S.; Huang, C.; Aivazian, G.; Ross, J. S.; Cobden, D. H.; Xu, X. Vapor-Solid Growth of High Optical Quality MoS2 Monolayers with near-Unity Valley Polarization. *ACS Nano* **2013**, *7* (3), 2768–2772. https://doi.org/10.1021/nn4002038.

(3) Li, H.; Wu, J.; Huang, X.; Lu, G.; Yang, J.; Lu, X.; Xiong, Q.; Zhang, H. Rapid and Reliable Thickness Identification of Two-Dimensional Nanosheets Using Optical Microscopy. *ACS Nano* **2013**, *7* (11), 10344–10353. https://doi.org/10.1021/NN4047474.

(4) Li, F.; Huang, T. De; Lan, Y. W.; Lu, T. H.; Shen, T.; Simbulan, K. B.; Qi, J. Anomalous Lattice Vibrations of CVD-Grown Monolayer MoS2 Probed Using Linear Polarized Excitation Light. *Nanoscale* **2019**, *11* (29), 13725–13730. https://doi.org/10.1039/C9NR03203G.

(5) Mak, K. F.; He, K.; Lee, C.; Lee, G. H.; Hone, J.; Heinz, T. F.; Shan, J. Tightly Bound Trions in Monolayer MoS2. *Nat. Mater. 2012 123* **2012**, *12* (3), 207–211. https://doi.org/10.1038/nmat3505.

(6) Kaniyoor, A.; Ramaprabhu, S. A Raman Spectroscopic Investigation of Graphite Oxide Derived Graphene. *AIP Adv.* **2012**, *2* (3), 032183. https://doi.org/10.1063/1.4756995.

(7) Perumbilavil, S.; Sankar, P.; Priya Rose, T.; Philip, R. White Light Z-Scan Measurements of Ultrafast Optical Nonlinearity in Reduced Graphene Oxide Nanosheets in the 700 Nm Region. *Appl. Phys. Lett.* **2015**, *107* (5), 051104.

https://doi.org/10.1063/1.4928124.

(8) Fichter, W. B. Some Solutions for the Large Deflections of Uniformly Loaded Circular Membranes. *NASA Tech. Pap.* **1997**, *3658*, 1–24.

(9) Lloyd, D.; Liu, X.; Boddeti, N.; Cantley, L.; Long, R.; Dunn, M. L.; Bunch, J. S. Adhesion, Stiffness, and Instability in Atomically Thin MoS2 Bubbles. *Nano Lett.* **2017**, *17* (9), 5329–5334. https://doi.org/10.1021/acs.nanolett.7b01735.

(10) Cooper, R. C.; Lee, C.; Marianetti, C. A.; Wei, X.; Hone, J.; Kysar, J. W. Nonlinear Elastic Behavior of Two-Dimensional Molybdenum Disulfide. *Phys. Rev. B - Condens. Matter Mater. Phys.* **2013**, *87* (3), 035423. https://doi.org/10.1103/PhysRevB.87.035423.

(11) Kang, J.; Sahin, H.; Peeters, F. M. Mechanical Properties of Monolayer Sulphides: A Comparative Study between MoS2, HfS2 and TiS3. *Phys. Chem. Chem. Phys.* **2015**, *17* (41), 27742–27749. https://doi.org/10.1039/C5CP04576B.

(12) Yang, R.; Lee, J.; Ghosh, S.; Tang, H.; Sankaran, R. M.; Zorman, C. A.; Feng, P. X. L. Tuning Optical Signatures of Single- and Few-Layer MoS2 by Blown-Bubble Bulge Straining up to Fracture. *Nano Lett.* **2017**, *17* (8), 4568–4575. https://doi.org/10.1021/ACS.NANOLETT.7B00730.

(13) Christopher, J. W.; Vutukuru, M.; Lloyd, D.; Bunch, J. S.; Goldberg, B. B.; Bishop, D. J.; Swan, A. K. Monolayer MoS2 Strained to 1.3% with a Microelectromechanical System. *J. Microelectromechanical Syst.* **2019**, *28* (2), 254–263. https://doi.org/10.1109/JMEMS.2018.2877983.

(14) Boddeti, N. G.; Koenig, S. P.; Long, R.; Xiao, J.; Bunch, J. S.; Dunn, M. L. Mechanics of Adhered, Pressurized Graphene Blisters. *J. Appl. Mech. Trans. ASME* **2013**, *80* (4). https://doi.org/10.1115/1.4024255.

(15) Wan, K. T.; Mai, Y. W. Fracture Mechanics of a New Blister Test with Stable Crack Growth. *Acta Metall. Mater.* **1995**, *43* (11), 4109–4115. https://doi.org/10.1016/0956-7151(95)00108-8.

(16) Dai, Z.; Rao, Y.; Lu, N. Two-Dimensional Crystals on Adhesive Substrates Subjected to Uniform Transverse Pressure. *Int. J. Solids Struct.* **2022**, *257*, 111829. https://doi.org/10.1016/J.IJSOLSTR.2022.111829.

(17) Suk, J. W.; Na, S. R.; Stromberg, R. J.; Stauffer, D.; Lee, J.; Ruoff, R. S.; Liechti, K. M. Probing the Adhesion Interactions of Graphene on Silicon Oxide by Nanoindentation. *Carbon N. Y.* **2016**, *103*, 63–72. https://doi.org/10.1016/J.CARBON.2016.02.079.

(18) Pickering, J. P.; Van Der Meer, D. W.; Vancso, G. J. Effects of Contact Time, Humidity, and Surface Roughness on the Adhesion Hysteresis of Polydimethylsiloxane. *J. Adhes. Sci. Technol.* **2012**, *15* (12), 1429–1441. https://doi.org/10.1163/156856101753213286.

(19) Greenwood, J. A.; Johnson, K. L. The Mechanics of Adhesion of Viscoelastic Solids. *Mater. Sci. Eng. R Reports* **2006**, *43* (3), 697–711. https://doi.org/10.1080/01418618108240402.

(20) Horn, R. G.; Israelachvili, J. N.; Pribac, F. Measurement of the Deformation and Adhesion of Solids in Contact. *J. Colloid Interface Sci.* **1987**, *115* (2), 480–492. https://doi.org/10.1016/0021-9797(87)90065-8.

(21) Kesari, H.; Doll, J. C.; Pruitt, B. L.; Cai, W.; Lew, A. J. Role of Surface Roughness in Hysteresis during Adhesive Elastic Contact. *Philos. Mag. Lett.* **2010**, *90* (12), 891–902. https://doi.org/10.1080/09500839.2010.521204.

(22) Schwartz, L. W.; Garoff, S. Contact Angle Hysteresis on Heterogeneous Surfaces. *Langmuir* **1985**, *1* (2), 219–230. https://doi.org/10.1021/LA00062A007.